%% file: HIN-16-010_temp.tex
\pdfoutput=1

\documentclass[11pt,twoside,a4paper,cmspaper,final,collab]{cms-tdr}
\def\svnVersion{378986}\def\cmsCernNoTag{CERN-EP/2016-147}\def\cmsCernDate{\today}\def\cmsMessage{Published in Physics Letters B as \href{http://dx.doi.org/10.1016/j.physletb.2016.12.009}{\doi{10.1016/j.physletb.2016.12.009.}}}

\begin{document}\cmsNoteHeader{HIN-16-010}

\hyphenation{had-ron-i-za-tion}
\hyphenation{cal-or-i-me-ter}
\hyphenation{de-vices}
\RCS$Revision: 373780 $
\RCS$HeadURL: svn+ssh://svn.cern.ch/reps/tdr2/papers/HIN-16-010/trunk/HIN-16-010.tex $
\RCS$Id: HIN-16-010.tex 373780 2016-11-14 20:14:13Z zhchen $
\newlength\cmsFigWidth
\ifthenelse{\boolean{cms@external}}{\setlength\cmsFigWidth{0.98\columnwidth}}{\setlength\cmsFigWidth{0.65\textwidth}}
\ifthenelse{\boolean{cms@external}}{\providecommand{\cmsLeft}{top\xspace}}{\providecommand{\cmsLeft}{left\xspace}}
\ifthenelse{\boolean{cms@external}}{\providecommand{\cmsRight}{bottom\xspace}}{\providecommand{\cmsRight}{right\xspace}}
\newcommand{\NA}{\text{---}\xspace}
\newcommand{\roots}{\ensuremath{\sqrt{s}}\xspace}
\newcommand{\rootsNN}{\ensuremath{\sqrt{s_{_\mathrm{NN}}}}\xspace}
\newcommand{\deta}{\ensuremath{\Delta\eta}\xspace}
\newcommand{\dphi}{\ensuremath{\Delta\phi}\xspace}
\newcommand{\pttrg}{\ensuremath{\pt^{\text{trig}}}\xspace}
\newcommand{\ptass}{\ensuremath{\pt^{\text{assoc}}}\xspace}
\newcommand{\ptref}{\ensuremath{\pt^{\text{ref}}}\xspace}
\newcommand{\vsecsig}{\ensuremath{v_2^\text{sig}}\xspace}
\newcommand{\vsecbkg}{\ensuremath{v_2^\text{bkg}}\xspace}
\newcommand{\vsecobs}{\ensuremath{v_2^\text{obs}}\xspace}
\newcommand{\ket}{\ensuremath{KE_{\mathrm{T}}}\xspace}
\newcommand{\pp}{\ensuremath{\Pp\Pp}\xspace}
\newcommand{\PbPb}{\ensuremath{\mathrm{PbPb}}\xspace}
\newcommand{\pPb}{\ensuremath{\Pp\mathrm{Pb}}\xspace}
\newcommand{\AonA}{\ensuremath{\mathrm{AA}}\xspace}
\newcommand{\dAu}{\ensuremath{\mathrm{dAu}}\xspace}
\newcommand{\noff}{\ensuremath{N_\text{trk}^\text{offline}}\xspace}
\providecommand{\PKzS}{\ensuremath{\mathrm{K^0_S}}}
\providecommand{\PgL}{\ensuremath{\Lambda}}
\providecommand{\PagL}{\ensuremath{\overline{\Lambda}}}
\providecommand{\PYTHIA}{\textsc{pythia}\xspace}
\providecommand{\Pap}{\ensuremath{\mathrm{\overline{p}}}}
\newcommand{\dmean}[1]{\langle\langle#1\rangle\rangle}
\newcommand{\cn}[1]{c_n\{#1\}}
\newcommand{\vn}[1]{\ensuremath{v_n\{#1\}}}
\cmsNoteHeader{HIN-16-010}
\title{Evidence for collectivity in pp collisions at the LHC}

\date{\today}

\abstract{
Measurements of two- and multi-particle angular correlations in pp collisions
at $\sqrt{s} = 5$, 7, and 13\TeV are presented as a function of charged-particle
multiplicity. The data, corresponding to integrated luminosities of
1.0\pbinv (5\TeV), 6.2\pbinv (7\TeV), and 0.7\pbinv (13\TeV), were collected using the CMS detector at the LHC.
The second-order ($v_2$) and third-order ($v_3$) azimuthal anisotropy harmonics of
unidentified charged particles, as well as $v_2$ of $\PKzS$ and $\PgL/\PagL$ particles,
are extracted from long-range two-particle correlations as functions of particle multiplicity
and transverse momentum.
For high-multiplicity pp events, a mass ordering is observed for
the $v_2$ values of charged hadrons (mostly pions), $\PKzS$, and $\PgL/\PagL$,
with lighter particle species exhibiting a stronger azimuthal anisotropy signal below $\pt \approx 2$\GeVc.
For 13\TeV data, the $v_2$ signals are also extracted from four- and six-particle correlations for the first time
in pp collisions, with comparable magnitude to those from two-particle correlations.
These observations are similar to those seen in pPb and PbPb collisions, and support the interpretation of a collective origin for the observed
long-range correlations in high-multiplicity pp collisions.

}

\hypersetup{%
pdfauthor={CMS Collaboration},%
pdftitle={Evidence for collectivity in pp collisions at the LHC},%
pdfsubject={CMS},%
pdfkeywords={CMS, physics, heavy ion, ridge, correlation, pp}}

\maketitle
\section{Introduction}
The observation of long-range two-particle azimuthal correlations at large
relative pseudorapidity ($\abs{\deta}$) in high final-state particle multiplicity (high-multiplicity) proton-proton
(\pp)~\cite{Khachatryan:2010gv,Aad:2015gqa,Khachatryan:2015lva}
and proton-lead (\pPb)~\cite{CMS:2012qk,alice:2012qe,Aad:2012gla,Aaij:2015qcq}
collisions at the LHC has opened up new opportunities for studying novel dynamics of particle
production in small, high-density quantum chromodynamic (QCD) systems.
A key feature of such correlations is an enhanced structure on the near side
(relative azimuthal angle $\abs{\dphi} \approx 0$) of two-particle $\deta$-$\dphi$ correlation functions that
extends over a wide range in relative pseudorapidity ($\abs{\deta} \approx 4$).
Such a long-range, near side correlation structure, known as the ``ridge'',
was first observed in relativistic nucleus-nucleus (\AonA) collisions from RHIC to LHC energies,
including copper-copper~\cite{Alver:2008gk}, gold-gold~\cite{Adams:2005ph,Abelev:2009af,Alver:2008gk,Alver:2009id,Abelev:2009jv},
and lead-lead (\PbPb)~\cite{Chatrchyan:2011eka,Chatrchyan:2012wg,Aamodt:2010pa,ATLAS:2012at,Chatrchyan:2012ta,CMS:2013bza} systems.
Based on extensive studies, it has been suggested that the hydrodynamic collective flow
of a strongly interacting and expanding medium~\cite{Ollitrault:1992bk,Heinz:2013th,Gale:2013da}
is responsible for these long-range correlations in these large heavy ion systems.
In hydrodynamic models, the detailed azimuthal correlation structure of emitted particles is
typically characterized by its Fourier components~\cite{Voloshin:1994mz}. In particular,
the second and third Fourier components, known as elliptic ($v_2$) and triangular ($v_3$) flow,
most directly reflect the medium response to, respectively, the initial collision geometry and
its fluctuations~\cite{Alver:2010gr}, providing insight into fundamental transport
properties of the medium~\cite{Alver:2010dn,Schenke:2010rr,Qiu:2011hf}.
Recently, at RHIC, such long-range correlations have also been observed and studied in lighter
\AonA systems such as \dAu \cite{Adare:2014keg,Adamczyk:2015xjc} and $^{3}$HeAu~\cite{Adare:2015ctn}.

In systems such as \pp and \pPb, where the transverse size of the overlap region is comparable to that of a single proton,
the formation of a hot and dense fluid-like medium was not expected.
The expectations for other small systems like \dAu and $^{3}$HeAu were similar.
Various theoretical models have been proposed to interpret the origin of the observed
long-range correlations in small collision systems~\cite{Li:2012hc,Bjorken:2013boa,Dusling:2015gta,Dusling:2012wy,Dusling:2012cg,Schenke:2014zha,Bozek:2011if,Bozek:2012gr}.
These include initial-state gluon correlations without final-state
interactions~\cite{Dusling:2012wy,Dusling:2012cg} or, similar to what is thought
to occur in \AonA systems, hydrodynamic flow that develops in a conjectured
high-density medium~\cite{Schenke:2014zha,Bozek:2011if,Bozek:2012gr}.

Owing to the magnitude of the correlation signal, significant progress has been made in unraveling
the nature of the ridge correlations in high-multiplicity \pPb collisions.
Measurements of anisotropy Fourier harmonics ($v_n$), using identified particles~\cite{ABELEV:2013wsa,Khachatryan:2014jra}
and multi-particle correlation techniques~\cite{Khachatryan:2015waa,Abelev:2014mda,Aad:2013fja,Chatrchyan:2013nka},
reveal features that support a collective
origin of the observed correlations~\cite{Werner:2013ipa,Bozek:2013ska,Yan:2013laa,Schenke:2016lrs}.

In high-multiplicity \pp collisions, the nature of the observed long-range correlation still remains
poorly understood.
Long-range correlations in pp collisions were first observed at $\roots = 7$\TeV~\cite{Khachatryan:2010gv},
and the extraction of anisotropy $v_2$ harmonics in \pp collisions was recently reported using data at $\roots = 13$\TeV~\cite{Aad:2015gqa}.
A better understanding of the underlying particle correlation mechanisms
leading to these observations requires more detailed study of the properties of the $v_2$
and higher-order harmonics in \pp collisions.
In particular, their dependence on particle species, and other aspects related to their possible collective nature, are
the key to scrutinize various theoretical interpretations. Furthermore, a quantitative
modeling of long-range correlations in \pPb collisions
is found to be highly sensitive to the spatial structure of the proton~\cite{Schenke:2014zha}.
Fluctuations of the substructure of nucleons are not well understood, but they can
be better constrained with studies of anisotropy harmonics in \pp collisions.

This paper extends the characterization of long-range correlation phenomena in
high-multiplicity \pp collisions by presenting a detailed study of
two- and multi-particle azimuthal correlations with unidentified charged particles, as well as correlations of
reconstructed $\PKzS$ and $\PgL/\PagL$ particles at various LHC collision energies.
The results of $v_{2}$ and $v_{3}$ harmonics, extracted from two-particle correlations,
are studied as functions of particle \pt and event multiplicity.
The residual contribution to long-range correlations of back-to-back jet correlations is estimated and removed
by subtracting correlations obtained from very low multiplicity \pp events.
The $v_2$ harmonics are also extracted using the
multi-particle cumulant method to shed light on the possible collective nature
of the correlations. The pp results are directly compared to those found for
\pPb and \PbPb  systems over a broad range of similar multiplicities.

\section{The CMS detector and data sets}

The central feature of the CMS apparatus is a superconducting solenoid
of 6\unit{m} internal diameter. Within the solenoid
volume, there are a silicon pixel and strip tracker detector, a lead tungstate crystal
electromagnetic calorimeter (ECAL), and a brass and scintillator hadron
calorimeter (HCAL), each composed of a barrel and two endcap sections.
Muons are measured in gas-ionization detectors embedded in the steel
flux-return yoke outside the solenoid.
The silicon tracker measures charged particles within the pseudorapidity
range $\abs{\eta}< 2.5$. It consists of 1440 silicon pixel and 15\,148
silicon strip detector modules and is located in the 3.8\unit{T} field of the solenoid.
For non-isolated particles of $1 < \pt < 10\GeVc$
and $\abs{\eta} < 1.4$, the track resolutions are typically 1.5\% in \pt and 25--90 (45--150)\mum in the transverse (longitudinal) impact parameter~\cite{TRK-11-001}.
Iron and quartz-fiber Cherenkov hadron forward (HF) calorimeters cover the range
$2.9 < \abs{\eta} < 5.2$ on either side of the interaction region.
These HF calorimeters are azimuthally subdivided into $20^{\circ}$ modular
wedges and further segmented to form $0.175{\times}0.175\unit{rad}^{2}$
$(\Delta\eta{\times}\Delta\phi)$ ``towers''.
The beam pickup for timing (BPTX) devices are designed to
provide precise information on the LHC bunch structure and timing of the incoming beams.
They are located around the beam pipe at a distance of 175\unit{m} from the interaction point on either
side. A more detailed description of the CMS detector, together with a definition
of the coordinate system used and the relevant kinematic variables,
can be found in Ref.~\cite{Chatrchyan:2008zzk}.
The detailed Monte Carlo (MC) simulation
of the CMS detector response is based on \GEANTfour~\cite{GEANT4}.

The data samples of pp collisions used in this analysis were collected by the CMS experiment
in 2010 at $\roots = 7$\TeV, and in 2015 at 5.02\TeV (labeled as 5\TeV for simplicity) and 13\TeV,
with integrated luminosities of 6.2, 1.0, and 0.7\pbinv,
respectively.

\section{Event and track selection}

Minimum bias (MB) \pp events were triggered by requiring
the coincidence of signals from both BPTX devices, indicating the presence
of two proton bunches crossing at the interaction point (zero bias condition).
The data used in this study were recorded with an average number
of pp interactions per bunch crossing ranging from 0.1 to 1.4.
Because of hardware limits on the data acquisition rate,
only a small fraction (${\sim}10^{-3}$) of all MB triggered events were recorded.
In order to collect a large sample of high-multiplicity \pp collisions, a dedicated
trigger was implemented using the CMS level-1 (L1) and high-level
trigger (HLT) systems. At L1, the total transverse energy
summed over ECAL and HCAL was required to be greater than a given threshold
(40, 45 and 55\GeV thresholds are used). Online track reconstruction for the HLT was based on
the three layers of pixel detectors, requiring the track origin to be located within a cylindrical region
centered on the nominal interaction point with a
length of 30\unit{cm} along the beam and a radius of 0.2\unit{cm} perpendicular to the beam.
For each event, the vertex reconstructed with the highest number of pixel tracks was selected.
The vertex reconstruction efficiency with high track multiplicities is 100\%.
The number of pixel tracks (${N}_\text{trk}^\text{online}$)
with $\abs{\eta}<2.4$, $\pt > 0.4\GeVc$, and a distance of closest approach less than 0.12\unit{cm}
to this vertex, was determined for each event. Data were taken with HLT thresholds of
${N}_\text{trk}^\text{online}\geq60,$ 85, 110, seeded with L1 total transverse energy thresholds of 40, 45, and 55\GeV, respectively.

In the offline analysis, hadronic collisions are selected by requiring
at least one HF calorimeter tower with more than 3\GeV of total energy in each
of the two HF detectors. Events are also required to contain at least one reconstructed
primary vertex within 15\unit{cm} of the nominal interaction point along the beam axis and
within 0.15\unit{cm} in the direction transverse to the beam trajectory. Beam related background
is suppressed by rejecting events for which less than 25\% of all reconstructed
tracks pass the \textit{high-purity} selection (as defined in Ref.~\cite{TRK-11-001}).
With these selection criteria, 94--96\% of the pp interactions simulated
with  \PYTHIA6 tune Z2~\cite{Sjostrand:2006za} and \PYTHIA8
tune CUETP8M1~\cite{Sjostrand:2007gs} event generators that have at least one
particle with energy $E>3$\GeV in both
ranges $-5<\eta <-3$ and $3<\eta <5$ are selected.

In this analysis, primary tracks, \ie tracks that emanate from the primary
vertex and that satisfy the high-purity criteria,
are used to define the event charged-particle multiplicity and perform correlation
measurements. Additional requirements are also applied to enhance the
purity of primary tracks. The significance of the separation along the beam
axis ($z$) between the track and the primary vertex, $d_z/\sigma(d_z)$, and the significance
of the impact parameter relative to the primary vertex transverse to the beam,
$d_\mathrm{T}/\sigma(d_\mathrm{T})$, must be smaller than 3. The relative uncertainty
of the transverse-momentum measurement, $\sigma(\pt)/\pt$, must be smaller than 10\%.
To ensure high tracking efficiency and to reduce the rate of misreconstructed
tracks, only tracks in the region $\abs{\eta}<2.4$ and $\pt > 0.3\GeVc$ are included.
The \pt threshold is raised to 0.4\GeVc\ for purposes of the event multiplicity
determination, to match the requirement in the HLT, while tracks down to
0.3\GeVc\ are used in the correlation analysis. Based on simulation studies
using {\sc geant4} to propagate particles from the \PYTHIA8\ event generator,
the combined geometrical acceptance and efficiency for primary
track reconstruction exceeds 60\% for $\pt \approx 0.3$\GeVc and $\abs{\eta}<2.4$.
The efficiency is greater than 90\% in the $\abs{\eta}<1$ region for $\pt>0.6\GeVc$.
For the event multiplicity range studied in this paper, no dependence of the
tracking efficiency on the multiplicity is found and the rate of misreconstructed tracks is 1--2\%.

Additionally, the CMS \textit{loose}~\cite{TRK-11-001} tracks are also
used to incorporate secondary-track candidates with larger track impact parameters,
for reconstructing $\PKzS$ and $\PgL/\PagL$ candidates (also called $V^{0}$ candidates).
The reconstruction of $V^{0}$ candidates in this analysis is identical to
that in Refs.~\cite{Khachatryan:2011tm,Khachatryan:2014jra}, where more details can be found.
Oppositely charged tracks that are detached
(having large $d_z/\sigma(d_z)$ and $d_\mathrm{T}/\sigma(d_\mathrm{T})$ values) from the primary vertex
are selected to determine if they point to a common secondary vertex. The pair of tracks are assumed to
be $\Pgpp\Pgpm$ in $\PKzS$ reconstruction, while the assumption of $\Pgpm\Pp (\Pgpp\Pap)$
is used in $\PgL$ ($\PagL$) reconstruction. The angle $\theta^{\text{point}}$ between the $V^0$ momentum vector and
the vector connecting the primary and $V^0$ vertices
is required to satisfy $\cos\left(\theta^{\text{point}}\right)>0.999$.
The $\PKzS$ (\PgL/\PagL) reconstruction efficiency is about 6\%
(1\%) for $\pt \approx 1\GeVc$ and 17\%\,(7\%) for $\pt>3\GeVc$ within $\abs{\eta}<2.4$.  This
efficiency includes the effects of acceptance and the branching ratio for $V^0$ particle decays into
neutral particles. The relatively low reconstruction efficiency of the $V^0$ candidates is primarily due
to the low efficiency for reconstructing daughter tracks with $\pt < 0.3$\GeVc\ or larger impact parameters.

Following the same procedure as used in previous analyses~\cite{Khachatryan:2010gv,CMS:2012qk,Chatrchyan:2013nka},
the \pp data sets at different energies are divided into classes of reconstructed
track multiplicity, \noff, where primary tracks with $\abs{\eta}<2.4$ and $\pt >0.4$\GeVc\
are counted. Details of the multiplicity classification in this analysis, including
the fractional inelastic cross section and the average number of primary tracks before and
after correcting for detector effects in each multiplicity range,
are provided in Table~\ref{tab:newmultbinning}.
Within a given \noff range, the average
event multiplicity is expected to be larger for higher \roots data, as suggested by
the average uncorrected \noff values. However, due to a slightly higher tracking efficiency,
and hence a smaller efficiency correction, the corrected average multiplicity,
$N_\text{trk}^\text{corrected}$, for the same \noff range happens to be smaller
for 7\TeV data than for 5\TeV data.
Uncertainties on $N_\text{trk}^\text{corrected}$ values come from
systematic uncertainties on the tracking efficiency correction factor estimated from MC simulations.

\begin{table*}[ht]\renewcommand{\arraystretch}{1.2}\addtolength{\tabcolsep}{-1pt}
\centering
\topcaption{ \label{tab:newmultbinning} Fraction of MB triggered events after event selections
in each multiplicity bin, and the average multiplicity of reconstructed tracks per bin
with $\abs{\eta}<2.4$ and $\pt >0.4$\GeVc, before (\noff) and after
($N_\text{trk}^\text{corrected}$) efficiency correction, for pp data at
$\roots = 5$, 7, and 13\TeV.}
\begin{tabular}{ l | l | l | l | l | l | l | l | l | l}
\hline
\multirow{2}{*}{\noff} & \multicolumn{3}{c|}{Fraction} & \multicolumn{3}{c|}{$\left<N_\text{trk}^\mathrm{offline}\right>$} & \multicolumn{3}{c}{$\left<N_\text{trk}^\text{corrected}\right>$} \\\cline{2-10}
 & 5\TeV & 7\TeV & 13\TeV & 5\TeV & 7\TeV & 13\TeV & 5\TeV & 7\TeV & 13\TeV \\
\hline
MB & 1.0 & 1.0 & 1.0 & 13 & 15 & 16 & 16$\pm$1 & 17$\pm$1 & 19$\pm$1  \\
$[0, 10)$  & 0.48 & 0.44 & 0.43 & 4.8 & 4.8 & 4.8 & 5.8$\pm$0.3 & 5.5$\pm$0.2 & 5.9$\pm$0.3 \\
$[10, 20)$ & 0.29 & 0.28 & 0.26 & 14 & 14 & 14 & 17$\pm$1 & 16$\pm$1 & 17$\pm$1 \\
$[20, 30)$ & 0.14 & 0.15 & 0.15 & 24 & 24 & 24 & 28$\pm$1 & 28$\pm$1 & 30$\pm$1 \\
$[30, 40)$ & 0.06 & 0.08 & 0.08 & 34 & 34 & 34 & 41$\pm$2 & 40$\pm$2 & 42$\pm$2 \\
$[40, 60)$ & 0.03 & 0.05 & 0.07 & 47 & 47 & 47 & 56$\pm$2 & 54$\pm$2 & 58$\pm$2 \\
$[60, 85)$ & $3\times10^{-3}$ & $7\times10^{-3}$ & 0.02 & 66 & 67 & 68 & 80$\pm$3 & 78$\pm$3 & 83$\pm$3 \\
$[85, 95)$ & $9\times10^{-5}$ & $3\times10^{-4}$ & $1\times10^{-3}$ & 88 & 89 & 89 & 106$\pm$4 & 103$\pm$4 & 109$\pm$4 \\
$[95, 105)$ & $2\times10^{-5}$ & $9\times10^{-5}$ & $5\times10^{-4}$ & 98 & 99 & 99 & 118$\pm$5 & 114$\pm$4 & 121$\pm$5 \\
$[105, 115)$ & $5\times10^{-6}$ & $2\times10^{-5}$ & $2\times10^{-4}$ & 108 & 109 & 109 & 130$\pm$5 & 126$\pm$5 & 133$\pm$5 \\
$[115, 125)$ & $1\times10^{-6}$ & $8\times10^{-6}$ & $6\times10^{-5}$ & 118 & 118 & 119 & 142$\pm$6  & 137$\pm$5 &145$\pm$6 \\
$[125, 135)$ & $2\times10^{-7}$ & $2\times10^{-6}$ & $2\times10^{-5}$ & 126 & 128 & 129 & 153$\pm$6 & 149$\pm$6 & 157$\pm$6  \\
$[135, 150)$ & $5\times10^{-8}$ & $4\times10^{-7}$ & $8\times10^{-6}$ & 139 & 140 & 140 & 167$\pm$7 & 162$\pm$6 & 171$\pm$7 \\
$[150, \infty)$ & $5\times10^{-9}$ & $8\times10^{-8}$ & $2\times10^{-6}$ & 155 & 156 & 158 & 186$\pm$8 & 181$\pm$7 & 193$\pm$8 \\
\hline
\end{tabular}
\end{table*}

\section{Analysis technique}

The analysis techniques for two- and multi-particle correlations are identical to those used in
Refs.~\cite{Chatrchyan:2011eka,Chatrchyan:2012wg,CMS:2012qk,Chatrchyan:2013nka,Khachatryan:2014jra,Khachatryan:2015lva,Khachatryan:2015waa}.
They are briefly summarized in this section for the analysis of the new \pp data samples.

\subsection{Two-particle correlations and Fourier harmonics}
\label{subsec:2pcorr}

For each track multiplicity class, ``trigger'' particles are defined as
charged particles or $V^{0}$ candidates with $\abs{\eta} < 2.4$ and originating from the primary vertex
within a given \pttrg\ range. The number of trigger particles in the event is
denoted by $N_\text{trig}$. Particle pairs are then formed by associating each
trigger particle with the remaining charged primary tracks with $\abs{\eta} < 2.4$ and from a specified
\ptass\ interval (which can be either the same as or different from the \pttrg\ range).
A pair is removed if the associated particle is the daughter of any trigger $V^{0}$ candidate
(this contribution is negligible since associated particles are mostly primary tracks).
The two-dimensional (2D) per-trigger-particle associated yield is defined in the same way
as done in previous analyses,
\begin{equation}
\label{2pcorr_incl}
\frac{1}{N_\text{trig}}\frac{\rd^{2}N^\text{pair}}{\rd\Delta\eta\, \rd\Delta\phi}
= B(0,0)\,\frac{S(\Delta\eta,\Delta\phi)}{B(\Delta\eta,\Delta\phi)},
\end{equation}
where $\Delta\eta$ and $\Delta\phi$ are the differences in $\eta$
and $\phi$ of the pair. The same-event pair distribution, $S(\Delta\eta,\Delta\phi)$,
represents the yield of particle pairs normalized by $N_\text{trig}$ from the same event,
\begin{equation}
\label{eq:signal}
S(\Delta\eta,\Delta\phi) = \frac{1}{N_\text{trig}}\frac{\rd^{2}N^\text{same}}{\rd\Delta\eta\, \rd\Delta\phi}.
\end{equation}
The mixed-event pair distribution,
\begin{equation}
\label{eq:background}
B(\Delta\eta,\Delta\phi) = \frac{1}{N_\text{trig}}\frac{\rd^{2}N^\text{mix}}{\rd\Delta\eta\, \rd\Delta\phi},
\end{equation}
is constructed by pairing the trigger particles in each
event with the associated charged particles from 20 different randomly selected
events in the same 0.5\unit{cm} wide $z_\mathrm{vtx}$ range and from the same track multiplicity class.
The same-event and mixed-event pair distributions are first calculated for each event, and
then averaged over all the events within the track multiplicity class.
The ratio $B(0,0)/B(\Delta\eta,\Delta\phi)$ mainly accounts for pair acceptance
effects, with $B(0,0)$ representing the mixed-event associated yield for
both particles of the pair going in approximately the same direction and
thus having maximum pair acceptance.
Following the procedure
described in Refs.~\cite{Chatrchyan:2011eka,Chatrchyan:2012wg,CMS:2012qk,Chatrchyan:2013nka,Khachatryan:2014jra},
each reconstructed primary track and $V^{0}$ candidate
is weighted by a correction factor derived from MC simulations, which accounts for detector effects including the
reconstruction efficiency, the detector acceptance, and the fraction of misreconstructed tracks.

The azimuthal anisotropy harmonics of charged particles, $\PKzS$ and \PgL/\PagL\ particles can be
extracted via a Fourier decomposition of long-range two-particle \dphi correlation functions,
obtained by averaging the 2D two-particle correlation function over $\abs{\deta}>2$
(to remove short-range correlations, such as those from jet fragmentation),
\begin{linenomath}
\begin{equation}
\label{eq:Vn}
\frac{1}{N_\text{trig}}\frac{\rd N^\text{pair}}{\rd\Delta\phi} = \frac{N_\text{assoc}}{2\pi} \left[ 1+\sum\limits_{n} 2V_{n\Delta} \cos (n\Delta\phi)\right],
\end{equation}
\end{linenomath}
where $V_{n\Delta}$ are the Fourier coefficients and $N_\text{assoc}$
represents the average number of pairs per trigger particle for a given
$(\pttrg, \ptass)$ bin. The first three Fourier terms are included
in the fits to the dihadron correlation functions. Including additional terms has a
negligible effect on the results of the Fourier fit.

Assuming $V_{n\Delta}$ coefficients can be factorized into the product of single-particle anisotropies~\cite{Chatrchyan:2013nka},
the elliptic and triangular
anisotropy harmonics, $v_{2}\{2,\abs{\deta}>2\}$ and $v_{3}\{2,\abs{\deta}>2\}$, of trigger
particles can be extracted as a function of \pt from the fitted Fourier coefficients from the two-particle correlation method,
\begin{linenomath}
\begin{equation}
\label{eq:Vnpt}
v_{n}(\pttrg) = \frac{V_{n\Delta}(\pttrg,\ptref)}{\sqrt{V_{n\Delta}(\ptref,\ptref)}},~~~~~~~~~~~~~ n=2, 3.
\end{equation}
\end{linenomath}
Here, a fixed \ptref\ range for the ``reference'' charged primary
particles is chosen to be $0.3<\pt<3.0$\GeVc\ to minimize correlations from jets
at higher \pt.

To extract $v_2$ signal for $V^{0}$ candidates, the invariant mass distributions are
fitted by a sum of two Gaussian functions with a common mean to describe the true $V^{0}$ signal peak,
and a fourth-order polynomial function to model the background, as done in Ref.~\cite{Khachatryan:2014jra}.
The $v_2$ values are first extracted for $V^0$ candidates from the peak region
(which contains contributions from genuine $V^0$, as well as background $V^0$ candidates from random combinatorics) and from a sideband region (which
contains only background $V^0$s from random combinatorics),
denoted as \vsecobs\ and \vsecbkg. Here the peak region is defined as the mass window of ${\pm}2\sigma$ around the center
of the $V^{0}$ candidate mass peak, where $\sigma$ is found from
the addition in quadrature of the standard deviations of the two Gaussian functions weighted by their yields.
The sideband region is defined as the mass window outside
${\pm}3\sigma$ mass range around the $V^{0}$ candidate mass peak to upper limit of 0.565\,(1.135)\GeV and lower limit of 0.430\,(1.155)\GeV for $\PKzS$ (\PgL/\PagL) particles. The $v_2$ signal for $V^{0}$ candidates can
then be calculated as
\begin{linenomath}
\begin{equation}
\label{eq:v2sig}
\vsecsig\ = \frac{\vsecobs-(1-f^\text{sig}) \, \vsecbkg }{f^\text{sig}},
\end{equation}
\end{linenomath}
where $f^\text{sig}$ represents the signal yield fraction in the peak region determined
from the fits to the mass distribution. This fraction
exceeds 80\% for \PgL/\PagL\ candidates with $\pt>1$\GeVc and is above 95\% for \PKzS\ candidates over
the entire \pt\ range.

Although a requirement of $\abs{\deta}>2$ can largely exclude near side
jet-like correlations for $v_{n}\{2\}$ extraction, contributions from back-to-back (\ie dijet)
correlations are still present in the long-range, away side ($\Delta\phi \approx \pi$) region, especially for \pp collisions. By assuming that the shape of the jet-induced correlations is invariant with event
multiplicity, a procedure of removing jet-like correlations in \pPb collisions
was proposed in Refs.~\cite{alice:2012qe,Aad:2012gla}.
The method consists of subtracting the results for low-multiplicity events, where the
ridge signal is not present, from those for high-multiplicity events.
For this analysis, a very similar low-multiplicity subtraction method developed for \pPb collisions~\cite{Chatrchyan:2013nka} is employed.
The Fourier
coefficients, $V_{n\Delta}$, extracted from Eq.~(\ref{eq:Vn}) for $10 \leq \noff < 20$
are subtracted from the $V_{n\Delta}$ coefficients extracted in the higher-multiplicity region, with
\ifthenelse{\boolean{cms@external}}{
\begin{linenomath}
\begin{equation}
\label{eq:vnsubperiph}
\begin{split}
V^\text{sub}_{n\Delta}=&V_{n\Delta}-V_{n\Delta}(10\leq\noff<20)\\
&\times\frac{N_\text{assoc}(10\leq\noff<20)}{N_\text{assoc}}\\
&\times\frac{Y_\text{jet}}{Y_\text{jet}(10\leq\noff<20)}.
\end{split}
\end{equation}
\end{linenomath}
}{
\begin{linenomath}
\begin{equation}
\label{eq:vnsubperiph}
V^\text{sub}_{n\Delta}=V_{n\Delta}-V_{n\Delta}(10\leq\noff<20)\,\frac{N_\text{assoc}(10\leq\noff<20)}{N_\text{assoc}}\,\frac{Y_\text{jet}}{Y_\text{jet}(10\leq\noff<20)}.
\end{equation}
\end{linenomath}
}
Here, $Y_\text{jet}$ represents the near side jet yield obtained by integrating
the difference of the short- and long-range event-normalized associated yields for each multiplicity class
as shown for $105 \leq \noff < 150$ in
Fig.~\ref{fig:Corr_Proj} (to be described in Section~\ref{subsec:CorrFcn}) over $\abs{\Delta\phi} < 1.2$.
The ratio, $Y_\text{jet}/Y_\text{jet}(10\leq\noff<20)$,
is introduced to account for the enhanced jet correlations resulting from the
selection of higher-multiplicity events. This jet subtraction procedure is verified using \PYTHIA6\ (Z2)
and \PYTHIA8 tune CUETP8M1 \pp simulations, where no jet modification from initial-
or final-state effects is present. The residual $V_{n\Delta}$ after subtraction is
found to be consistent with zero.
The azimuthal anisotropy harmonics $v_n$ after correcting for back-to-back jet correlations estimated from
low-multiplicity data (denoted as $v_{n}^\text{sub}$)
can be extracted from $V^\text{sub}_{n\Delta}$ using Eq.~(\ref{eq:Vnpt}) and~(\ref{eq:v2sig}).
In this paper, both the $v_{n}$ and $v_{n}^\text{sub}$ results are presented.

\subsection{Fourier harmonics from multi-particle correlations}

To avoid a model-dependent correction of jet-like correlations in extracting $v_n$
harmonics from two-particle correlations, a multi-particle cumulant analysis using
the Q-cumulant method~\cite{Bilandzic:2010jr} is employed to
determine the second-order elliptic harmonic,
similar to what was done in \pPb and \PbPb\ collisions~\cite{Chatrchyan:2013nka,Khachatryan:2015waa}.
By simultaneously correlating several (no less than four)
particles, the multi-particle cumulant technique has the advantage of suppressing
short-range two-particle correlations such as jets and resonance decays. It
also serves as a powerful tool to directly probe the collective nature of the
observed azimuthal correlations.

The two-, four-, and six-particle azimuthal correlations~\cite{Bilandzic:2010jr} are evaluated as:
\begin{equation}\begin{split}
\dmean{2} &\equiv
    \bigl<\!\bigl< \re^{in(\phi_{1} - \phi_{2})} \bigr>\!\bigr>,
\\
\dmean{4} &\equiv
    \bigl<\!\bigl< \re^{in(\phi_{1} + \phi_{2} - \phi_{3} - \phi_{4})} \bigr>\!\bigr>,
\\
\dmean{6} &\equiv
    \bigl<\!\bigl< \re^{in(\phi_{1} + \phi_{2} + \phi_{3} - \phi_{4} - \phi_{5}
    - \phi_{6})} \bigr>\!\bigr>.
\label{eq:corr}
\end{split}\end{equation}
Here $\phi_{i}$ $(i=1,\ldots,6)$ are the
azimuthal angles of one unique combination of multiple particles in an event, $n$ is the harmonic number,
and $\bigl\langle\bigl\langle \cdots \bigr\rangle\bigr\rangle$ represents
the average over all combinations from all events within a given multiplicity range.
The corresponding cumulants, $\cn{4}$ and $\cn{6}$, are calculated as follows~\cite{Bilandzic:2010jr}:
\begin{equation}
\begin{split}
\cn{4} =& \dmean{4} - 2 \times \dmean{2}^2,
\\
\cn{6} =& \dmean{6} - 9 \times \dmean{4}\dmean{2} + 12 \times \dmean{2}^3.
\label{eq:cn}
\end{split}
\end{equation}

The Fourier harmonics $v_n$ that characterize the global azimuthal behavior
are related to the multi-particle cumulants~\cite{Bilandzic:2010jr} using
\begin{equation}
\begin{split}
\vn{4} &= \sqrt[4]{-\cn{4}},
\\
\vn{6} &= \sqrt[6]{\frac{1}{4} \cn{6}}.
\label{eq:v24}
\end{split}
\end{equation}
Note that a non-imaginary $\vn{4}$ coefficient, which would suggest a bulk medium collective behavior,
requires having a negative $\cn{4}$ value.

\subsection{Systematic uncertainties}

Systematic uncertainties in this analysis include those from the experimental procedure
for obtaining the two-particle $v_n$ harmonics,
as well as from the jet subtraction procedure.

The experimental systematic effects are evaluated by varying conditions in
extracting $v_{2}^\text{sub}\{2\}$, $v_{3}^\text{sub}\{2\}$, $v_{2}\{4\}$ and $v_{2}\{6\}$ values. The systematic uncertainties are found to have
no significant dependence on \pt and \roots so they are quoted to be constant
percentages over the entire \pt range for all collision energies.
Experimental systematic uncertainties due to track quality requirements are
examined by varying the track selection thresholds for $d_z/\sigma(d_z)$ and $d_{xy}/\sigma(d_{xy})$
from 2 to 5. A comparison of high-multiplicity \pp data for a given multiplicity
range but collected by two different HLT triggers with different trigger
efficiencies is made to study potential trigger biases. The possible contamination of
residual pileup events, especially for 5 and 7\TeV data, is also investigated
by varying the pileup selection of events in performing the analysis, from
no pileup rejection at all to selecting events with only one reconstructed vertex.
The sensitivity of the results to the primary vertex position ($z_\mathrm{vtx}$) is quantified by comparing results at
different $z_\mathrm{vtx}$ locations over a 30\unit{cm} wide range.

In the jet subtraction procedure for $v_{n}\{2\}$ measurements, while
the factor $Y_\text{jet}/Y_\text{jet}(10\leq\noff<20)$ accounts for any bias in the magnitude
of jet-like associated yield due to multiplicity selection, a change in the \dphi width of
away side yields could lead to residual jet effects in $v_{n}\{2\}$ results. This systematic
uncertainty is evaluated by integrating the associated yields in the $\abs{\Delta\eta} > 2$ region over
fixed \dphi windows of $\abs{\dphi} < \pi/3$ and $\abs{\dphi-\pi} < \pi/3$ on the near and away sides, respectively.
When extracting $v_{n}^\text{sub}$ results, 
the $Y_\text{jet}$ parameter in Eq.~(\ref{eq:vnsubperiph}) is then replaced by this difference of the near and away side yields.
By taking the difference of the yields in two \dphi windows symmetric around $\dphi=\pi/2$,
contributions from the second and fourth Fourier components are cancelled. By choosing the
\dphi window size to be $2\pi/3$, any contribution from the third Fourier component to the
near and away side associated yields is also cancelled. Any dependence of this yield difference on the event
multiplicity (beyond that induced by the $Y_\text{jet}/Y_\text{jet}(10\leq\noff<20)$ factor)
would indicate a modification of jet correlation width in \dphi.
The systematic uncertainty of $v_n$
due to this effect is estimated to be 16\%, 9\%, and 6\% for $\noff<40$, $40\leq\noff<85$, and $\noff>85$, respectively.
In the same sense, any multiplicity dependence of the \deta distribution of the away side would indicate a modification of the jet correlation.
The \deta distribution is investigated in a fixed window $\abs{\dphi-\pi} <\pi/16$ for different \noff ranges, resulting in systematic uncertainties of
8\%, 3\%, and 2.5\% for $\noff<40$, $40\leq\noff<85$, and $\noff>85$, respectively.
In addition, by separating events in a given multiplicity range into two groups corresponding to the top and bottom
30\% in the leading particle \pt distribution, jet correlations are either strongly
enhanced or suppressed in a controlled manner. After applying the subtraction procedure,
the $v_n$ results for the two event groups are consistent within 5\%.

Table~\ref{tab:syst-table3} summarizes various sources of systematic uncertainties
in this analysis for multiplicity-dependent results. The same sources apply to \pt differential results,
leading to total experimental systematic uncertainty of 5\% and uncertainties from the jet subtraction procedure of 9\%, 13\%, 23\%, and 37\%
for $\pttrg<2.2$\GeVc, $2.2 \leq \pttrg<3.6$\GeVc, $3.6 \leq \pttrg<4.6$\GeVc, and $\pttrg \geq 4.6$\GeVc, respectively.

\begin{table*}[ht]
\topcaption{\label{tab:syst-table3} Summary of systematic uncertainties
for multiplicity-dependent $v_{n}^\text{sub}\{2\}$ from two-particle correlations (after correcting for jet
correlations), and $v_{2}\{4\}$, $v_{2}\{6\}$ from multi-particle correlations in \pp collisions.
Different multiplicity ranges are represented as $[m,n)$.}
\centering
\begin{tabular}{l|c|c|c|c|c|c|c|c}
\hline
\multirow{2}*{Source} & \multicolumn{3}{c|}{$v_{2}^\text{sub}\{2\}$ (\%)} & \multicolumn{3}{c|}{$v_{3}^\text{sub}\{2\}$ (\%)} & \multicolumn{2}{c}{$v_{2}\{4\}$, $v_{2}\{6\}$ (\%)} \\
\cline{2-9}
          & [0,40) & [40,85) & [85,$\infty$) & [0,40) & [40,85) & [85,$\infty$) & [0,85) & [85,$\infty$) \\
\hline
 HLT trigger bias		&  \NA  & \NA  & 2 &  \NA  & \NA  & 2 & \NA & 2\\
 Track quality cuts		&  1 & 1 & 1 &  1 & 1 & 1 & 1 & 1 \\
 Pileup effects		    &  1.5 & 1.5 & 1.5 &  1.5 & 1.5 & 1.5 & 1.5 & 1.5 \\
 Vertex dependence		&  1.5 & 1.5 & 1.5 &  1.5 & 1.5 & 1.5 & 1.5 & 1.5 \\
 Jet subtraction		&  18 & 9.5 & 6.5 &  26.8 & 17 & 8.5	 & \NA & \NA \\
\hline
 Total		&	18.2 & 9.8 & 7.2	&	27 & 17.3 & 8.8	& 2.4	& 3.1 \\
\hline
\end{tabular}
\end{table*}

Systematic uncertainties originating from different sources are added in quadrature
to obtain the overall systematic uncertainty shown as boxes in the figures.
No energy dependence has been observed for the systematic uncertainties, therefore, they are only shown for 13 \TeV results.
Because of insufficient statistical precision, the uncertainties in $v_3$ resulting from the experimental procedure are assumed to be the same as those in $v_2$,
as was done in Refs.~\cite{Chatrchyan:2013nka,Khachatryan:2014jra}.
For the same reason, the systematic uncertainties
on the $v_{2}\{2\}$ results for $V^{0}$ particles that result from the variation of selection criteria,
alternative detector geometry and a MC closure test are obtained from studies performed for \pPb\ collisions in Ref.~\cite{Khachatryan:2014jra}, while those resulting from systemtic bias of the HLT trigger and jet subtraction method are taken to be the same as for the inclusive charged particles.
Different particle species have different $\eta$ distributions, which can affect the comparison of results
if there is a strong $\eta$ dependence.
This effect is found to be negligible
by comparing $v_{2}\{2\}$ results for $V^{0}$ particles with different reconstruction efficiency corrections for the $\eta$ distribution.
The relative systematic uncertainties for the
two-particle $V_{n\Delta}$ coefficients as a function of \noff in Fig.~\ref{fig:C2C3} (described in Section~\ref{subsec:vn2})
are exactly twice those for the corresponding $v_{n}$ harmonics, since $V_{n\Delta} = v_{n}^{2}$
when trigger and associated particles are selected from the same \pt range.
In the same way, relative systematic uncertainties for multi-particle $c_{2}\{m\}$ measurements
as a function of \noff in Fig.~\ref{fig:C4} (described in Section~\ref{subsec:cumu}) are exactly $m$ times those for the corresponding
$v_{2}\{m\}$ harmonics, where $m=4$ or 6.

\section{Results}

\subsection{Two-particle correlation functions}
\label{subsec:CorrFcn}

\begin{figure*}[thbp]
  \begin{center}
    \includegraphics[width=0.39\textwidth]{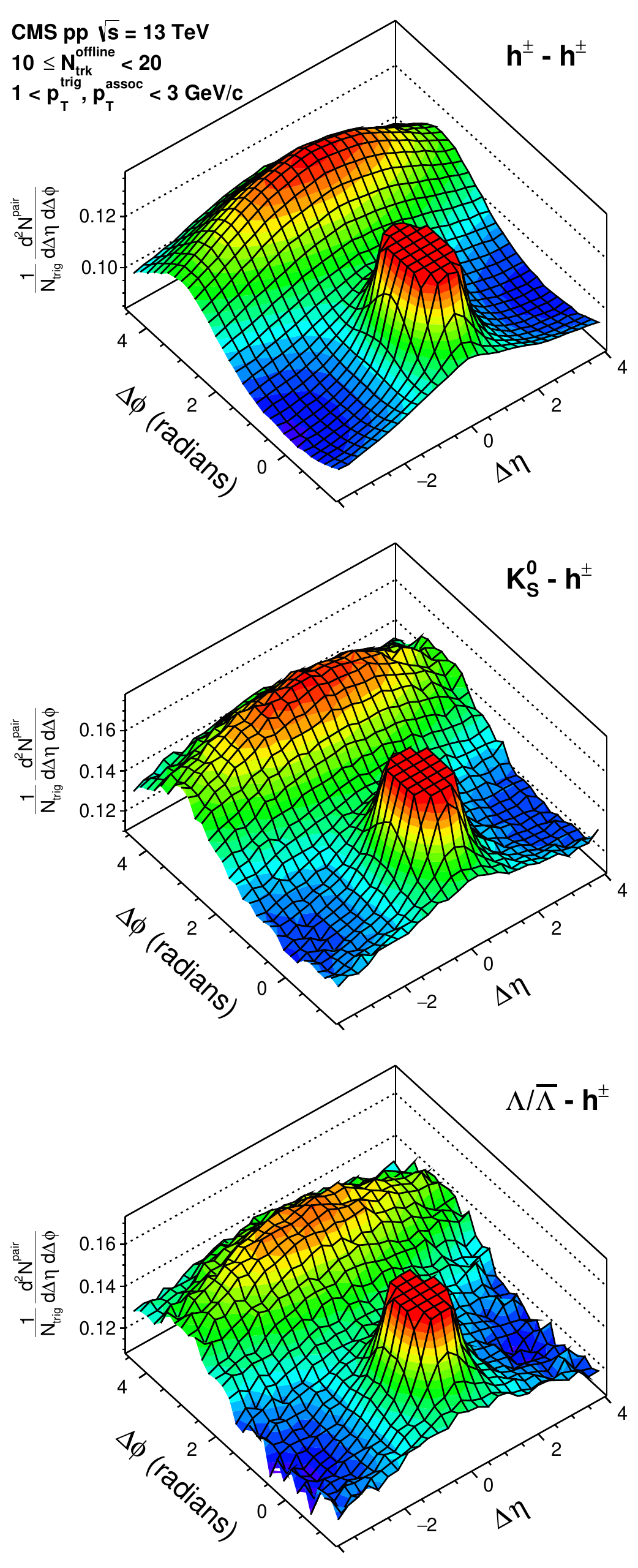}
    \includegraphics[width=0.39\textwidth]{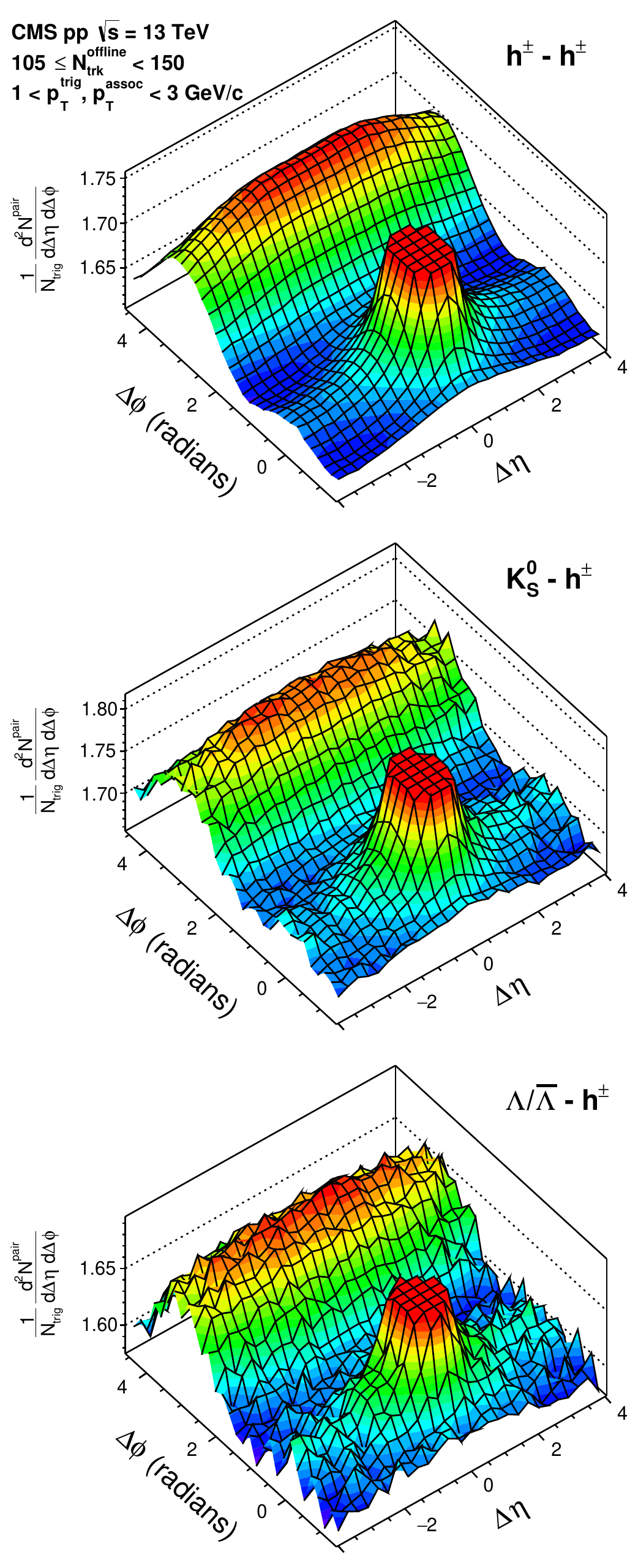}
    \caption{ The 2D two-particle correlation functions for inclusive charged
    particles (top), $\PKzS$ particles (middle), and \PgL/\PagL\ particles (bottom),
    with $1<\pttrg<3\GeVc$ and associated charged particles with $1<\ptass<3\GeVc$,
    in low-multiplicity ($10 \leq \noff < 20$, left) and high-multiplicity
    ($105 \leq \noff < 150$, right) \pp collisions at $\roots = 13$\TeV.
    }
    \label{fig:corr2D}
  \end{center}
\end{figure*}

Figure~\ref{fig:corr2D} shows the 2D \deta--\dphi correlation functions,
for pairs of a charged (top), a $\PKzS$ (middle), or a \PgL/\PagL\ (bottom)
trigger particle with a charged associated particle, in low-multiplicity ($10 \leq \noff < 20$, left)
and high-multiplicity ($105 \leq \noff < 150$, right) \pp collisions at $\roots = 13$\TeV.
Both trigger and associated particles are selected from the \pt range of 1--3\GeVc.
For all three types of particles at high multiplicity, in addition to the correlation
peak near $(\Delta\eta, \Delta\phi) = (0, 0)$ that results from jet fragmentation,
a long-range ridge structure is seen at $\dphi \approx 0$ extending at least 4 units
in $\abs{\deta}$, while such a structure is not observed in low multiplicity events.
On the away side ($\Delta\phi \approx \pi$) of the correlation functions, a long-range
structure is also seen and found to exhibit a much larger magnitude compared to that on the
near side for this \pt range. This away side correlation structure contains
contributions from back-to-back jets, which need to be accounted for before extracting any
other source of correlations.

\begin{figure*}[t!hb]
\centering
\includegraphics[width=\linewidth]{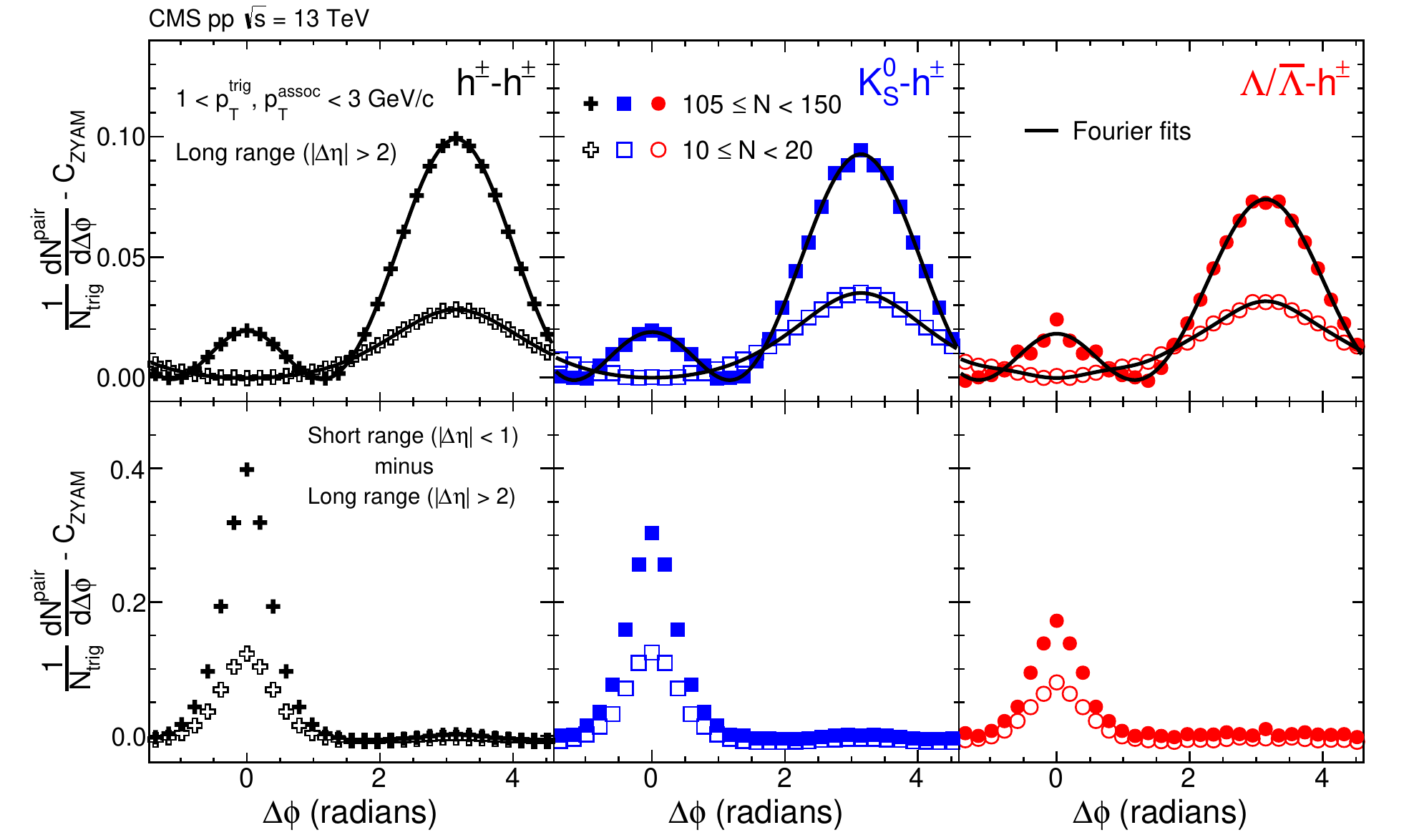}
  \caption{ \label{fig:Corr_Proj}
     The 1D \dphi correlation functions for the long-range (top) and short- minus long-range (bottom) regions
     after applying the ZYAM procedure
     in the multiplicity range $10 \leq \noff < 20$ (open symbols) and
     $105 \leq \noff < 150$ (filled symbols) of \pp collisions at $\roots = 13$\TeV,
     for trigger particles composed of inclusive charged particles (left, crosses), $\PKzS$ particles
     (middle, squares), and \PgL/\PagL\ particles (right, circles).
	A selection of 1--3\GeVc\ for both \pttrg\ and \ptass\ is used in all cases.
     }
\end{figure*}

To investigate the observed correlations in finer detail, the 2D distributions
shown in Fig.~\ref{fig:corr2D} are reduced to
one-dimensional (1D) distributions in $\Delta\phi$ by averaging over $\abs{\Delta\eta} < 1$
(defined as the ``short-range region") and $\abs{\Delta\eta} > 2$ (defined as the ``long-range region"),
respectively, as done in Refs.~\cite{CMS:2012qk,Khachatryan:2010gv,Chatrchyan:2011eka,Chatrchyan:2012wg}.
Figure~\ref{fig:Corr_Proj} shows examples of 1D \dphi correlation
functions for trigger particles composed of inclusive charged particles (left), $\PKzS$ particles (middle), and \PgL/\PagL\ particles (right), in the multiplicity range
$10 \leq \noff < 20$ (open symbols) and $105 \leq \noff < 150$ (filled symbols).
The curves show the Fourier fits from Eq.~(\ref{eq:Vn}) to the long-range region,
which will be discussed in detail in Section~\ref{subsec:vn2}.
To represent the correlated portion of the associated yield, each distribution is shifted to have zero
associated yield at its minimum following the standard zero-yield-at-minimum (ZYAM) procedure~\cite{Chatrchyan:2013nka}.
An enhanced correlation at
$\Delta\phi \approx 0$ in the long-range region is observed for $105 \leq \noff < 150$,
while such a structure is not presented for $10 \leq \noff < 20$. As illustrated in
Fig.~\ref{fig:corr2D} (right), the near side long-range ridge structure remains nearly
constant in $\Delta\eta$. Therefore, the near side jet correlation can be extracted by
taking a difference of 1D $\Delta\phi$ projections between the short- and long-range regions,
as shown in the bottom panels in Fig.~\ref{fig:Corr_Proj}.

\begin{figure*}[thb!p]
\centering
\includegraphics[width=\linewidth]{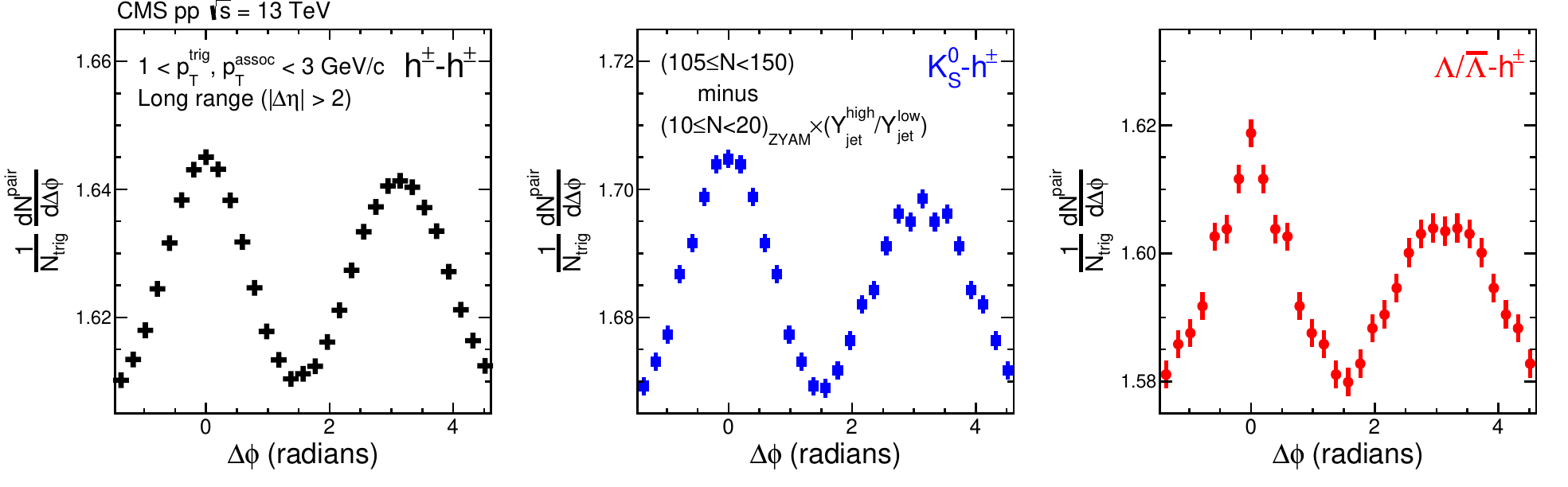}
  \caption{ \label{fig:Corr_Proj_sub}
     The 1D \dphi correlation functions for the long-range regions
     in the multiplicity range $105 \leq \noff < 150$
     of \pp collisions at $\roots = 13$\TeV, after subtracting scaled results from $10 \leq \noff <20$ with the ZYAM procedure applied.
     A selection of 1--3\GeVc\ for both \pttrg\ and \ptass\ is used in all cases.}
\end{figure*}

After subtracting the results, with the ZYAM procedure applied, for low-multiplicity $10 \leq \noff <20$
scaled by $Y_\text{jet}/Y_\text{jet}(10\leq\noff<20)$ as in Eq.~(\ref{eq:vnsubperiph}),
the long-range 1D \dphi correlation functions in the high-multiplicity range
$105 \leq \noff < 150$ for \pp collisions at $\roots = 13$\TeV are shown in
Fig.~\ref{fig:Corr_Proj_sub}, for trigger particles composed of inclusive
charged particles (left), $\PKzS$ (middle), and \PgL/\PagL\ (right) particles.
A ``double-ridge'' structure on the near and away side is observed after subtraction
of jet correlations. The shape of this structure, which is dominated by a second-order Fourier component,
is similar to what has been observed in \pPb~\cite{CMS:2012qk,alice:2012qe,Aad:2012gla,Aaij:2015qcq}
and \PbPb~\cite{Chatrchyan:2011eka,Chatrchyan:2012wg,Aamodt:2011by,ATLAS:2012at,Chatrchyan:2012ta} collisions.

\subsection{Two-particle fourier harmonics \texorpdfstring{$v_n$}{c[n]}}
\label{subsec:vn2}

\begin{figure*}[thbp]
\centering
\includegraphics[width=0.98\textwidth]{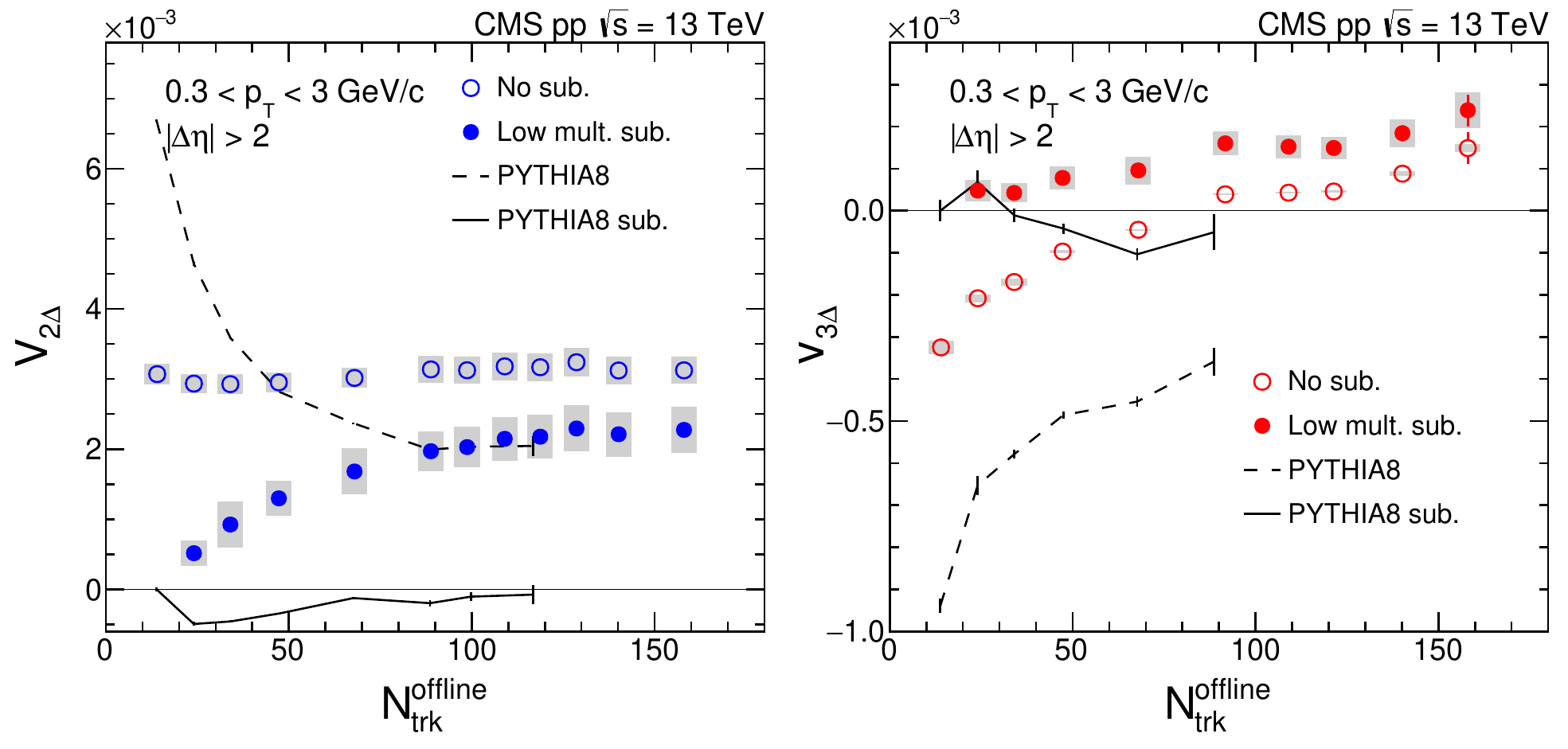}
  \caption{ \label{fig:C2C3}
     The second-order (left) and third-order (right) Fourier coefficients,
     $V_{2\Delta}$ and $V_{3\Delta}$, of
     long-range ($\abs{\deta}>2$) two-particle \dphi correlations
     as a function of \noff for charged particles, averaged over $0.3 < \pt < 3.0\GeVc$,
     in \pp collisions at $\roots = 13$\TeV, before (open) and after (filled)
     correcting for back-to-back jet correlations, estimated from the $10 \leq \noff < 20$ range.
     Results from \PYTHIA8\ tune CUETP8M1 simulation are shown as curves.
     The error bars correspond to statistical uncertainties, while the shaded areas denote the systematic uncertainties.
   }
\end{figure*}

Fourier coefficients, $V_{n\Delta}$, extracted from 1D $\Delta\phi$ two-particle correlation functions
for the long-range \deta region using Eq.~(\ref{eq:Vn}), are first studied for inclusive
charged hadrons. Figure~\ref{fig:C2C3} shows the $V_{2\Delta}$ and $V_{3\Delta}$ values for
pairs of inclusive charged particles averaged over $0.3  < \pt < 3.0\GeVc$ as a function of multiplicity
in \pp collisions at $\roots = 13$\TeV, before and after correcting for back-to-back jet
correlations estimated from low-multiplicity data ($10 \leq \noff < 20$).
The $V_{n\Delta}$ results for 5 and 7 \TeV are equal to the 13 \TeV results within the uncertainties.

Before corrections, the $V_{2\Delta}$ coefficients are found to remain relatively
constant as a function of multiplicity. This behavior is very different from the \PYTHIA8\ tune CUETP8M1
MC simulation, where the only source of long-range correlations is back-to-back jets and
the $V_{2\Delta}$ coefficients decrease with \noff. The $V_{3\Delta}$ coefficients found
using the \PYTHIA8\ simulation are always negative because of dominant contributions at $\dphi \approx \pi$
from back-to-back jets~\cite{Aamodt:2011by}, with their magnitudes decreasing as a function of \noff.
A similar trend is seen in the data for the
low multiplicity region, $\noff<90$. However, for $\noff\geq90$, the $V_{3\Delta}$ coefficients
in \pp data change to positive values. This transition directly indicates
a new phenomena that is not present in the \PYTHIA8\ simulation.
After applying the jet correction procedure detailed in Section~\ref{subsec:2pcorr},
$V_{2\Delta}$ exhibits an increase with multiplicity for $\noff \lesssim 100$, and
reaches a relatively constant value for the higher \noff region. The $V_{3\Delta}$ values
after subtraction of jet correlations become positive over the entire multiplicity range
and increase with multiplicity.

\begin{figure}[thb]
\centering
\includegraphics[width=\cmsFigWidth]{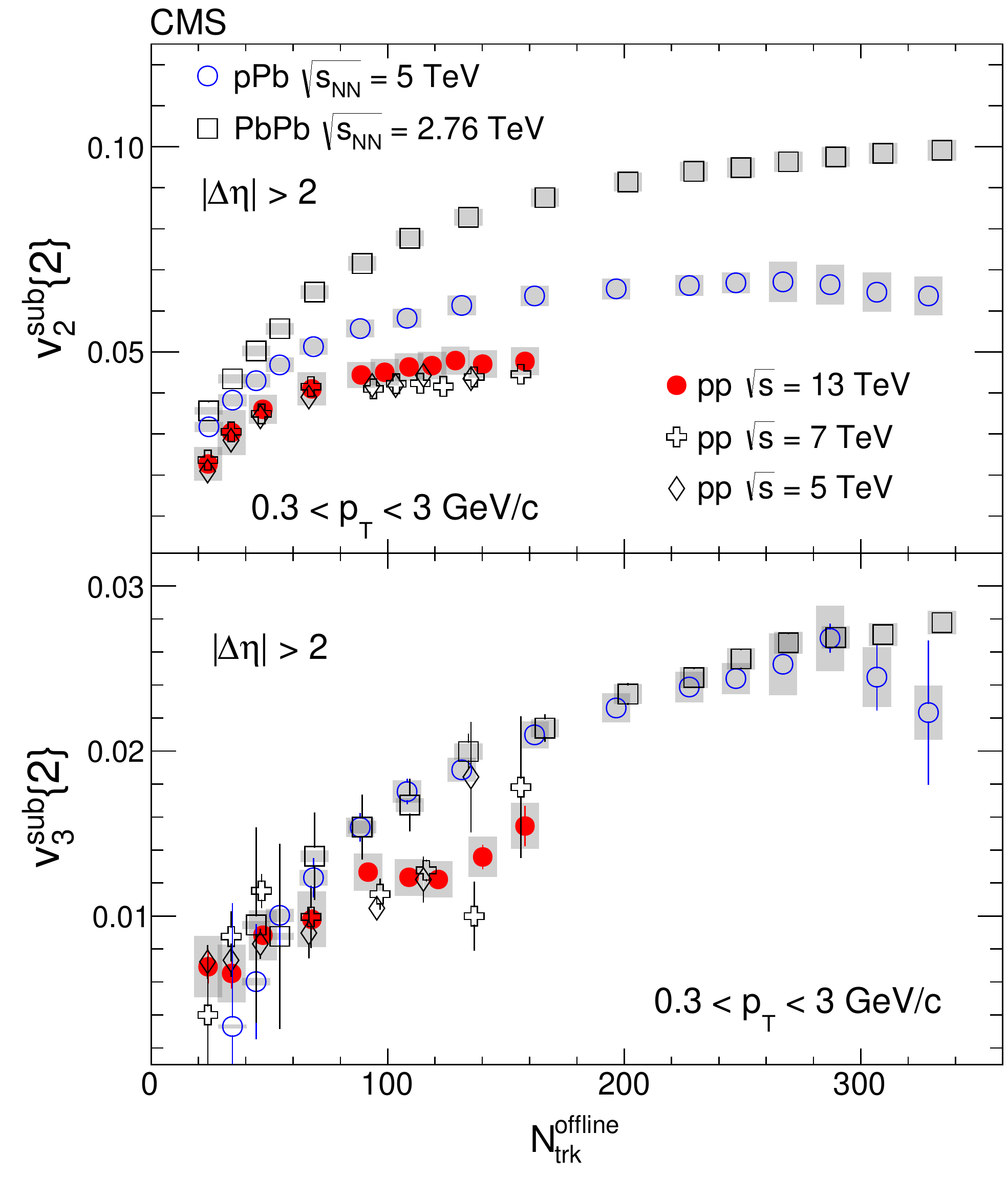}
  \caption{\label{fig:v2v3vsN}
     The $v_{2}^\text{sub}$ (top) and $v_{3}^\text{sub}$ (bottom) results of
     charged particles as a function of \noff, averaged over $0.3<\pt<3.0\GeVc$,
     in \pp collisions at $\roots = 5$, 7, and 13\TeV, \pPb collisions
     at $\rootsNN = 5$\TeV, and \PbPb\ collisions $\rootsNN = 2.76$\TeV,
     after correcting for back-to-back jet correlations estimated from low-multiplicity data.
     The error bars correspond to the statistical uncertainties, while the shaded areas denote the systematic uncertainties.
     Systematic uncertainties are found to have no dependence on \roots for pp results and therefore are only shown for 13\TeV.
   }
\end{figure}

The elliptic ($v_2$) and triangular ($v_3$) flow harmonics for charged particles
with $0.3 < \pt < 3.0\GeVc$, after applying the jet correction procedure, are then
extracted from the two-particle Fourier coefficients obtained using Eq.~(\ref{eq:Vnpt}),
and are shown in Fig.~\ref{fig:v2v3vsN} for \pp collisions at $\roots = 5$, 7, and 13\TeV.
The previously published \pPb data at $\rootsNN = 5$\TeV and \PbPb data at
$\rootsNN = 2.76$\TeV \cite{Chatrchyan:2013nka} are also shown for comparison among
different collision systems.

Within experimental uncertainties, for \pp collisions at all three
energies, there is no or only a very weak energy dependence for the $v_{2}^\text{sub}$ values.
The $v_{2}^\text{sub}$ results for \pp collisions
show a similar pattern as the \pPb results, becoming relatively constant as \noff increases,
while the \PbPb\ results show a moderate increase over
the entire \noff range shown in Fig.~\ref{fig:v2v3vsN}. Overall, the \pp data show a smaller $v_{2}^\text{sub}$ signal
than \pPb data over a wide multiplicity range, and both systems show smaller
$v_{2}^\text{sub}$ values than for the \PbPb\ system.

The $v_{3}^\text{sub}$ values of
the \pp data are comparable to those observed in \pPb and \PbPb\ collisions
in the very low multiplicity region $\noff < 60$, although
systematic uncertainties are large for all the three systems. At higher \noff,
$v_{3}^\text{sub}$ in \pp collisions
increases with multiplicity, although at a slower rate than observed in \pPb and \PbPb\ collisions.

The $v_2$ values reported by the ATLAS collaboration for \pp collisions
at $\roots = 13$\TeV in Ref.~\cite{Aad:2015gqa} have a different multiplicity dependence than
the results presented in this paper.
A nearly constant $v_2$ value is observed over the entire multiplicity range.
This distinct difference, especially in the low multiplicity region,
is rooted in the different approaches employed in identifying the $v_2$
signal from jet-like correlations. In the method from CMS and also
previous ATLAS and ALICE analyses~\cite{Aad:2012gla,Aad:2014lta,alice:2012qe},
$v_2$ is always extracted with respect to all of the particles in each event.
As seen in Eq.~(\ref{eq:Vn}), the Fourier coefficients in the current analysis represent an oscillation multiplied by the full $N_\text{assoc}$.
In the new approach by the ATLAS collaboration~\cite{Aad:2015gqa},
$v_2$ is extracted with respect to a subset of particles in each event.
In Ref.~\cite{Aad:2015gqa}, their equivalent of our Eq.~(\ref{eq:Vn}) uses a smaller number than the full $N_\text{assoc}$ in the events,
thereby assuming that some of the particles do not participate in the underlying processes producing the observed azimuthal correlations.
As a result, using the method of Ref.~\cite{Aad:2015gqa}, a $\cos (2\Delta\phi)$ modulation with exactly the same amplitude would yield a bigger
$V_{n\Delta}$ parameter compared to that found using Eq.~(\ref{eq:Vn}).
This, in turn, leads to larger $v_2$ values comparing to results obtained with
respect to all of the particles.
The difference between the two methods becomes larger as \noff decreases.
It was checked that when applying exactly the same kinematic selections
and analysis methods, no discrepancy is found between the
two experiments.
In the study of $v_2$ from multiparticle correlations,
as will be discussed in Section~\ref{subsec:cumu}, the $v_2$ is always considered
with respect to all the particles in the event
for each multiplicity class, which is consistent with the method used in this paper to extract $v_2$
from two-particle correlations.

\begin{figure}[thbp]
\centering
\includegraphics[width=0.49\textwidth]{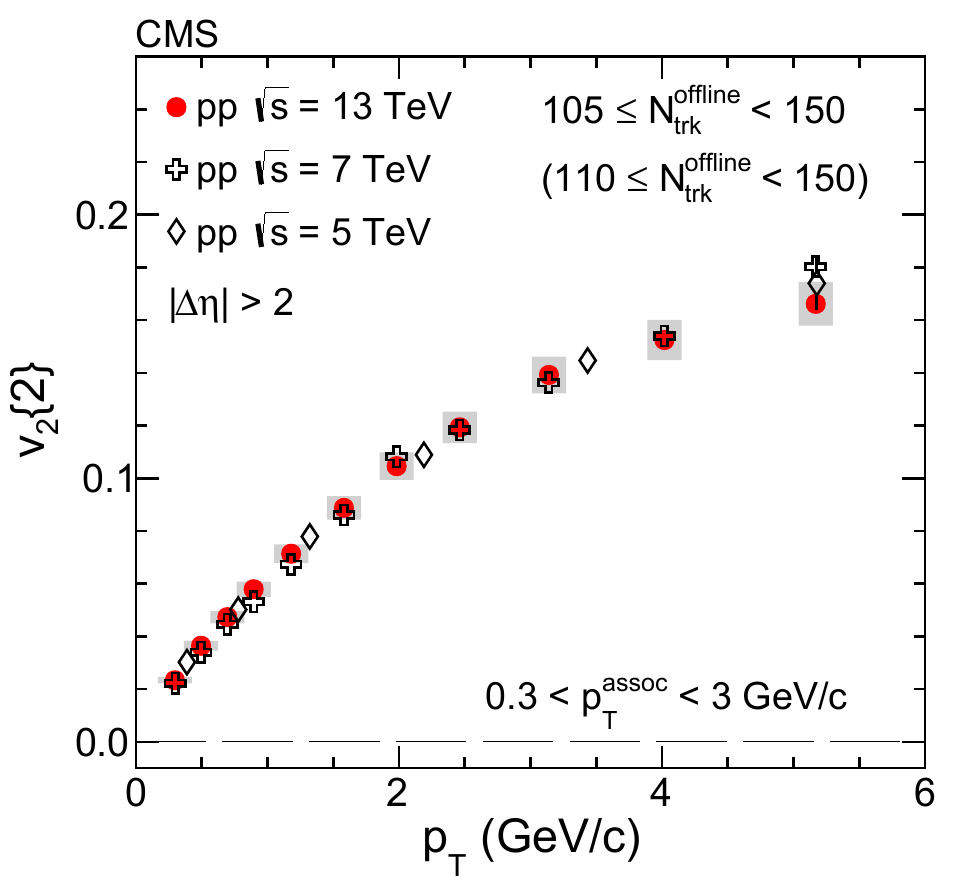}
\includegraphics[width=0.49\textwidth]{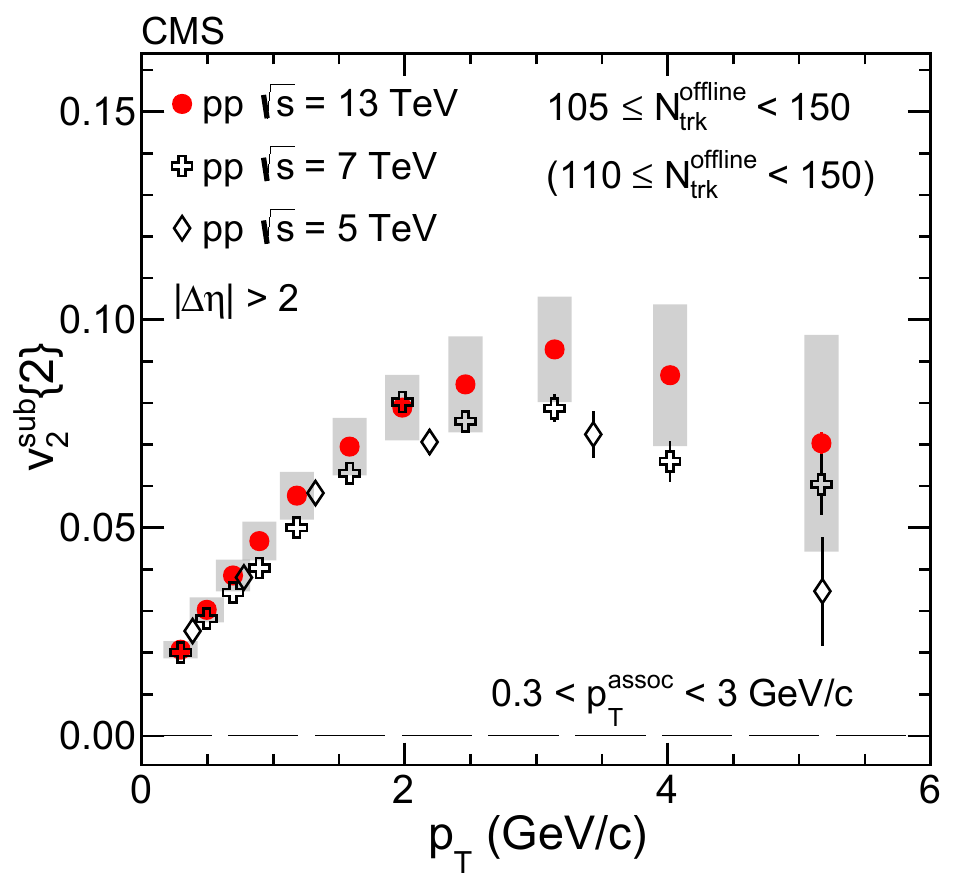}
  \caption{\label{fig:v2vspT}
  The $v_2$ results of inclusive charged particles, before (\cmsLeft) and after (\cmsRight)
  subtracting correlations from low-multiplicity events, as a function of \pt  in \pp collisions at $\roots = 13$\TeV for $105 \leq \noff < 150$ and at $\roots = 5$, 7\TeV for $110 \leq \noff < 150$.
 The error bars correspond to the statistical uncertainties, while the shaded areas denote the systematic uncertainties.
 Systematic uncertainties are found to have no dependence on \roots for pp results and therefore are only shown for 13\TeV.
   }
\end{figure}

The $v_2$ results as a function of \pt for high-multiplicity
pp events at $\roots = 5$, 7, and 13\TeV are shown in Fig.~\ref{fig:v2vspT}
before (\cmsLeft) and after (\cmsRight)
correcting for jet correlations.
To compare results with similar average \noff, $105 \leq \noff < 150$ is chosen for 13\TeV while $110 \leq \noff < 150$ is chosen for 5 and 7\TeV.
Little energy dependence is observed for the $\pt$-differential $v_{2}$ results,
especially before correcting for jet correlations, as shown in Fig.~\ref{fig:v2vspT} (\cmsLeft).
This conclusion also holds after jet correction procedure for $v_{2}^\text{sub}$ results (Fig.~\ref{fig:v2vspT}, \cmsRight)
within systematic uncertainties, although systematic uncertainties for $v_{2}^\text{sub}$ are
significantly higher at high \pt because of the large magnitude of the subtracted term.
This observation is consistent with the energy independence of associated long-range yields
on the near side reported in Ref.~\cite{Khachatryan:2015lva}.
The observed \pt dependence of $v_{2}^\text{sub}$, in high-multiplicity \pp events with peak values at 2--3\GeVc\
at various energies, is similar to that in \pPb~\cite{Chatrchyan:2013nka,Aad:2014lta,ABELEV:2013wsa} and \PbPb~\cite{Chatrchyan:2012wg,ATLAS:2011ah,Adam:2016izf} collisions.

\begin{figure}[thbp]
\centering
\includegraphics[width=0.49\textwidth]{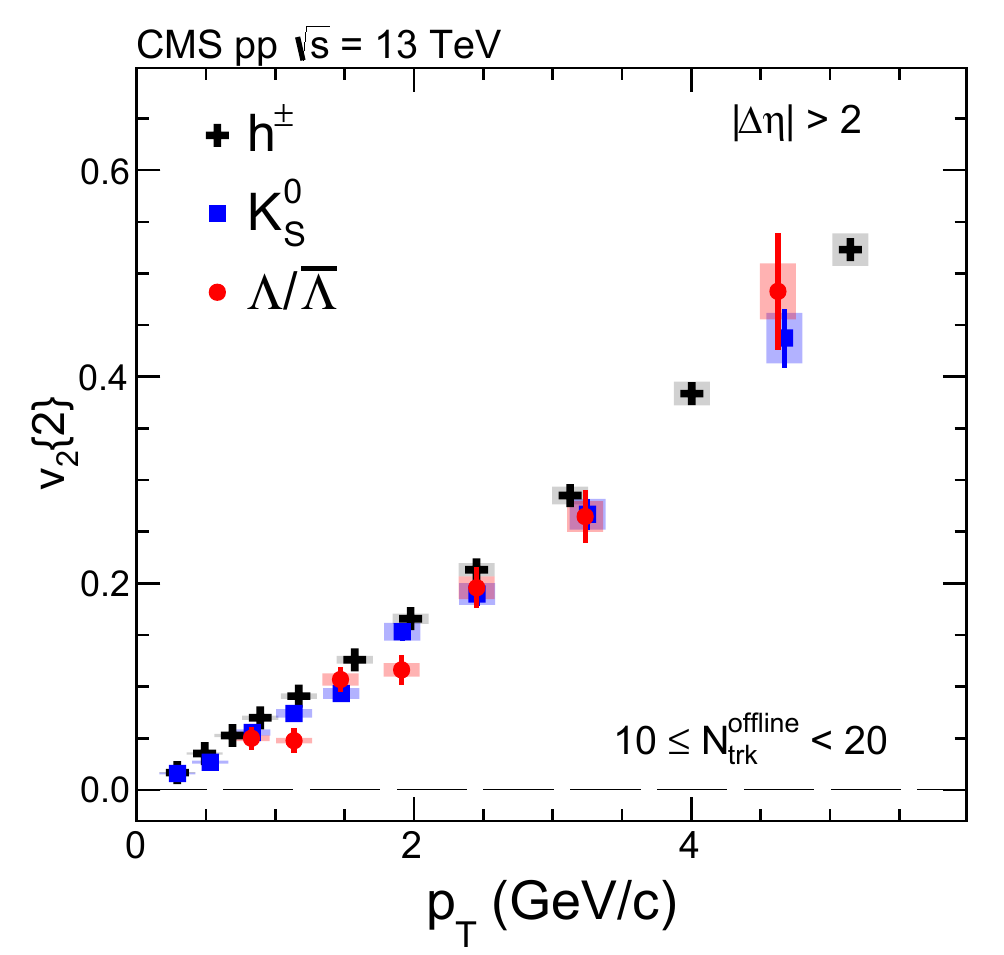}
\includegraphics[width=0.49\textwidth]{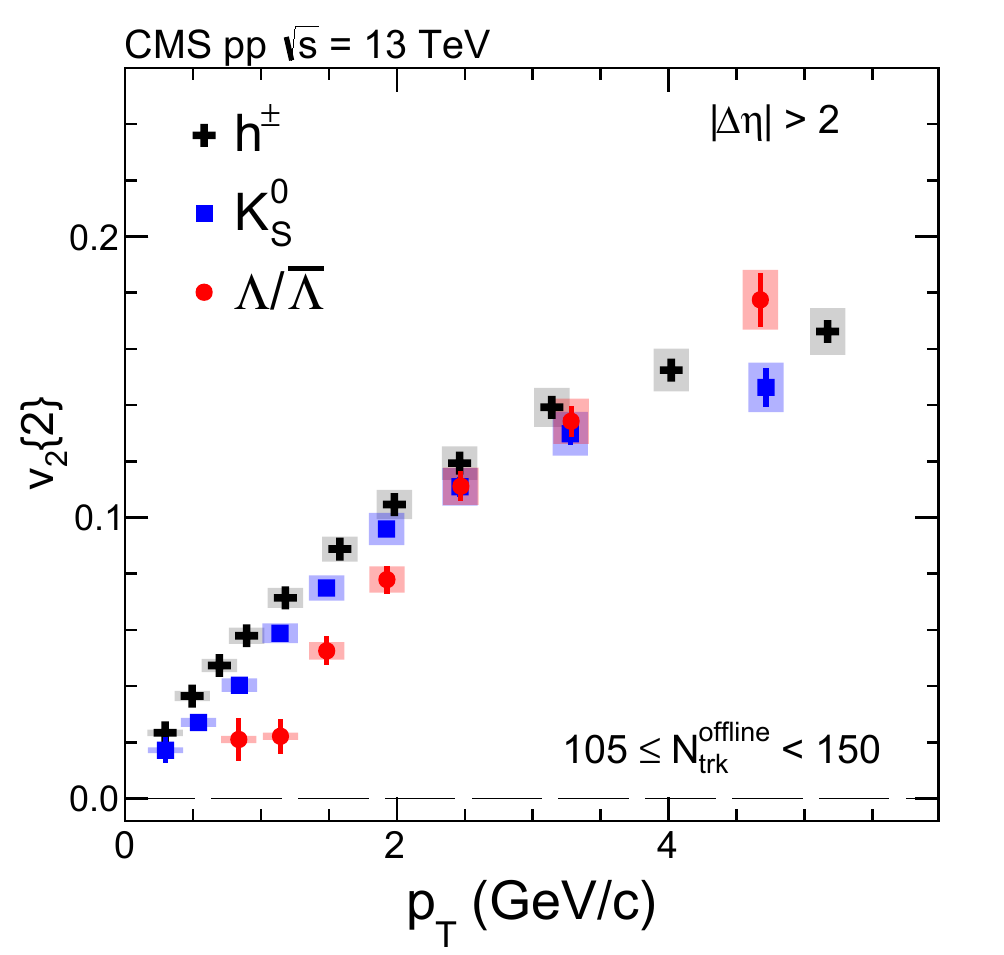}
  \caption{\label{fig:v2vspTPID}
     The $v_2$ results for inclusive charged particles, $\PKzS$ and \PgL/\PagL\ particles
     as a function of \pt in \pp collisions at $\roots = 13$\TeV,
     for $10 \leq \noff < 20$ (\cmsLeft) and $105 \leq \noff < 150$ (\cmsRight).
     The error bars correspond to the statistical uncertainties, while the shaded areas denote the systematic uncertainties.
   }
\end{figure}

\begin{figure}[thb]
\centering
\includegraphics[width=\cmsFigWidth]{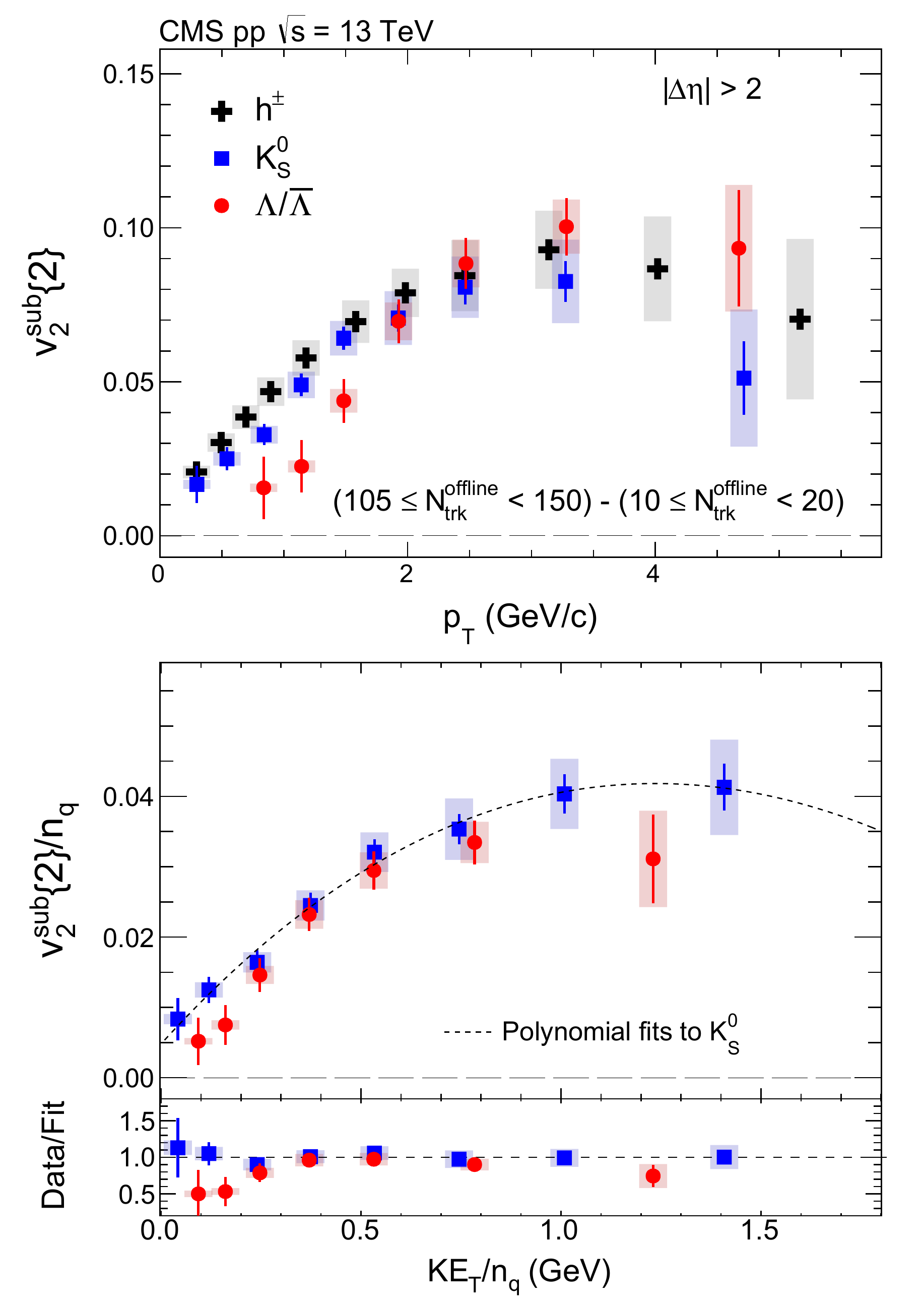}
  \caption{\label{fig:NCQsub}
     Top: the $v_{2}^\text{sub}$ results of inclusive charged particles, $\PKzS$ and
     \PgL/\PagL\ particles as a function of \pt for $105 \leq \noff < 150$,
     after correcting for back-to-back jet correlations
     estimated from low-multiplicity data. Bottom: the $n_\mathrm{q}$-scaled $v_{2}^\text{sub}$ results
     for $\PKzS$ and \PgL/\PagL\ particles as a function of $\ket/n_\mathrm{q}$.
     Ratios of $v_{2}^\text{sub}/n_\mathrm{q}$ for $\PKzS$ and \PgL/\PagL\ particles to
     a smooth fit function of data for $\PKzS$ particles are also shown.
     The error bars correspond to the statistical uncertainties, while the shaded areas denote the systematic uncertainties.
   }
\end{figure}

The dependence of the elliptic flow harmonic on particle species can shed further light on the nature of the correlations.
The $v_2$ data as a function of \pt for identified $\PKzS$ and \PgL/\PagL\
particles are extracted for \pp collisions at $\roots = 13$\TeV.
Figure~\ref{fig:v2vspTPID} shows the results for a low ($10 \leq \noff < 20$)
and a high ($105 \leq \noff < 150$) multiplicity range before applying the
jet correction procedure.

At low multiplicity (Fig.~\ref{fig:v2vspTPID} \cmsLeft), the $v_{2}$ values
are found to be similar for charged particles, $\PKzS$ and \PgL/\PagL\ hadrons
across most of the \pt range within statistical uncertainties, similar to the
observation in \pPb collisions at $\rootsNN = 5$\TeV~\cite{Khachatryan:2014jra}.
This would be consistent with the expectation that back-to-back jets are the
dominant source of long-range correlations on the away side in low-multiplicity
\pp events. Moving to high-multiplicity \pp events
($105 \leq \noff < 150$, Fig.~\ref{fig:v2vspTPID} \cmsRight),
a clear deviation of $v_{2}$ among various particle species
is observed. In the lower \pt region of $\lesssim 2.5\GeVc$,
the $v_{2}$ value of $\PKzS$ is greater than that of \PgL/\PagL\ at a given \pt value.
Both are consistently below the inclusive charged particle $v_{2}$ values.
Since most charged particles are pions in this \pt range, this indicates that
lighter particle species exhibit a stronger azimuthal anisotropy signal.
A similar trend was first observed in \AonA collisions at RHIC~\cite{Abelev:2007qg,Adare:2012vq},
and later also seen in \pPb collisions at the LHC~\cite{ABELEV:2013wsa,Khachatryan:2014jra}.
This behavior is found to be qualitatively consistent with both hydrodynamic
models~\cite{Werner:2013ipa,Bozek:2013ska} and an alternative initial state interpretation~\cite{Schenke:2016lrs}.
At $\pt>2.5$\GeVc, the $v_{2}$
values of \PgL/\PagL\ particles tend to become greater than those of $\PKzS$ particles.
This reversed ordering of $\PKzS$ and \PgL/\PagL\ at high \pt is similar
to what was previously observed in \pPb and \PbPb\ collisions~\cite{Khachatryan:2014jra}.

After applying the correction for jet correlations, the $v_{2}^\text{sub}$ results as a
function of \pt for $105 \leq \noff < 150$ are shown in Fig.~\ref{fig:NCQsub} (top) for the identified particles and charged hadrons.
The $v_{2}^\text{sub}$ values for all three types of particles are found to increase
with \pt, reaching 0.08--0.10 at $2<\pt<3$\GeVc, and then show a trend of decreasing $v_{2}^\text{sub}$ values for
higher \pt values. The particle mass ordering of $v_{2}$ values in the lower \pt region is also observed after applying jet correction procedure, while at higher \pt the ordering tends to reverse. As done in Ref.~\cite{Khachatryan:2014jra}, the scaling
behavior of $v_{2}^\text{sub}$ divided by the number of constituent quarks, $n_\mathrm{q}$, as a function
of transverse kinetic energy per quark, $\ket/n_\mathrm{q}$, is investigated for high-multiplicity
\pp events in Fig.~\ref{fig:NCQsub} (bottom). The dashed curve corresponds to a polynomial
fit to the $\PKzS$ data. The ratio of $n_\mathrm{q}$-scaled $v_{2}^\text{sub}$ results for $\PKzS$ and \PgL/\PagL\ particles divided by this polynomial function fit is also shown
in Fig.~\ref{fig:NCQsub} (bottom). An approximate scaling is seen for $\ket/n_\mathrm{q} \gtrsim 0.2 \GeV$
within about $\pm$10\%.

\subsection{Multi-particle correlations and collectivity}
\label{subsec:cumu}

\begin{figure}[thbp]
\centering
\includegraphics[width=0.495\textwidth]{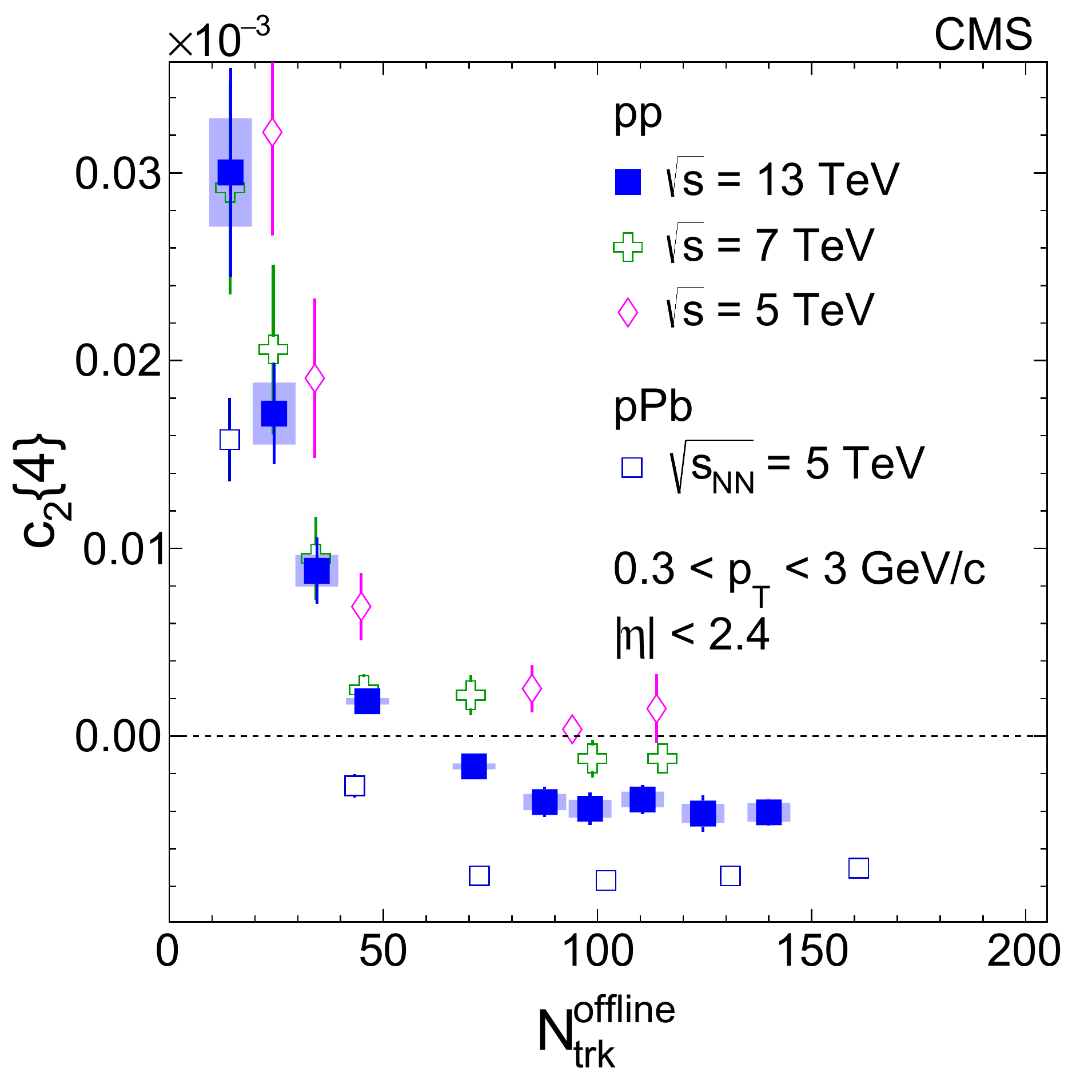}
\includegraphics[width=0.495\textwidth]{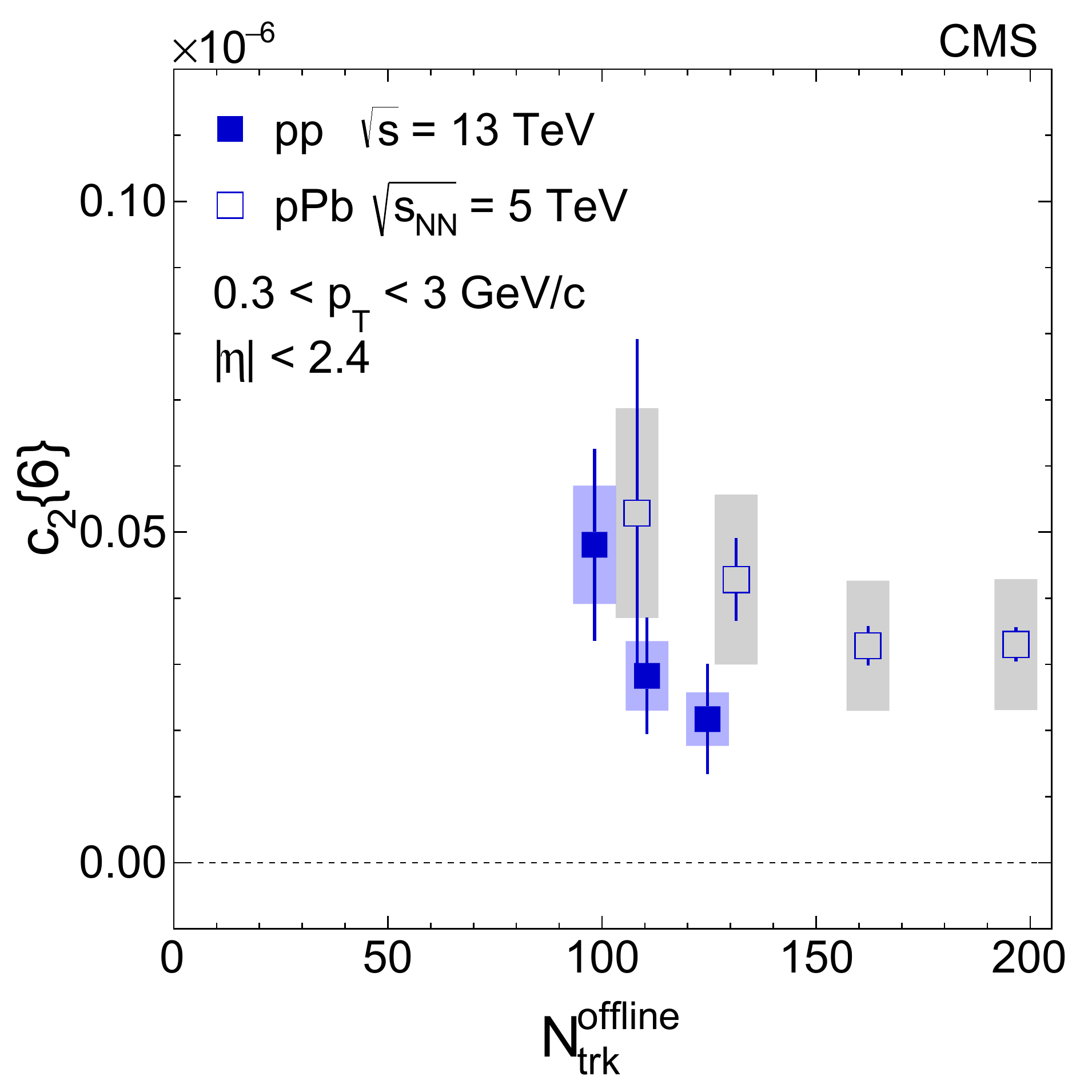}
  \caption{ \label{fig:C4}
     The $c_{2}\{4\}$ (\cmsLeft) and $c_{2}\{6\}$ (\cmsRight) values
     as a function of \noff for charged particles,
     averaged over $0.3 < \pt < 3.0\GeVc$ and $\abs{\eta}<2.4$, in \pp collisions at
     $\roots = 5$, 7, and 13\TeV. The \pPb data at $\rootsNN = 5$\TeV are
     also plotted for comparison.
     The error bars correspond to the statistical uncertainties, while the shaded areas denote the systematic uncertainties.
     Systematic uncertainties are found to have no dependence on \roots for pp results and therefore are only shown for 13\TeV.
   }
\end{figure}

To further reduce the residual jet correlations on the away side and explore the
possible collective nature of the long-range correlations, a four- and six-particle
cumulant analysis is used to extract the elliptic flow harmonics, $v_2\{4\}$
and $v_2\{6\}$. The four-particle cumulant $c_{2}\{4\}$ values for charged
particles with $0.3 < \pt < 3.0\GeVc$ are shown in Fig.~\ref{fig:C4} (\cmsLeft),
as a function of \noff for \pp collisions at $\roots = 5$, 7, and 13\TeV.
The \pPb data at $\rootsNN = 5$\TeV~\cite{Chatrchyan:2013nka} are also plotted
for comparison. The six-particle cumulant $c_{2}\{6\}$ values for \pp collisions
at $\roots = 13$\TeV are shown in Fig.~\ref{fig:C4} (\cmsRight), compared with \pPb data at $\rootsNN = 5$\TeV~\cite{Chatrchyan:2013nka}.
Due to statistical limitations, $c_{2}\{6\}$ values are only derived for
high multiplicities (\ie, $\noff \approx100$) for 13\TeV \pp data.

The $c_{2}\{4\}$ values for pp data at all energies show a
decreasing trend with increasing multiplicity, similar to that found for \pPb collisions.
An indication of energy dependence of $c_{2}\{4\}$ values is seen in Fig.~\ref{fig:C4}
(\cmsLeft), where $c_{2}\{4\}$ tends to be larger for a given \noff range
at lower \roots energies. As average \pt values are slightly smaller at lower collision energies,
the observed energy dependence may be related to smaller negative contribution to $c_{2}\{4\}$
from smaller $\pt$-averaged $v_2\{4\}$ signals.
In addition, when selecting from a fixed multiplicity range,
a larger positive contribution to $c_{2}\{4\}$ from larger jet-like correlations in the much rarer high-multiplicity events in lower energy \pp collisions
can also result in an energy dependence.
At \noff $\approx$ 60 for 13\TeV \pp data, the $c_{2}\{4\}$ values become and remain negative as the multiplicity increases further.
This behavior is similar to that observed for \pPb data where the sign change occurs at $\noff \approx 40$, indicating
a collective $v_2\{4\}$ signal~\cite{Borghini:2000sa}. For \pp data at $\roots = 5$ and 7\TeV, no significant
negative values of $c_{2}\{4\}$ are observed within statistical uncertainties.
The $c_{2}\{6\}$ values for pp data at 13\TeV show an increasing trend with decreasing multiplicity, similar to that found for \pPb collisions.
This trend might be due to a larger contribution to $c_{2}\{6\}$ from jet-like correlations in lower-multiplicity events.

\begin{figure*}[thb]
\centering
\includegraphics[width=\linewidth]{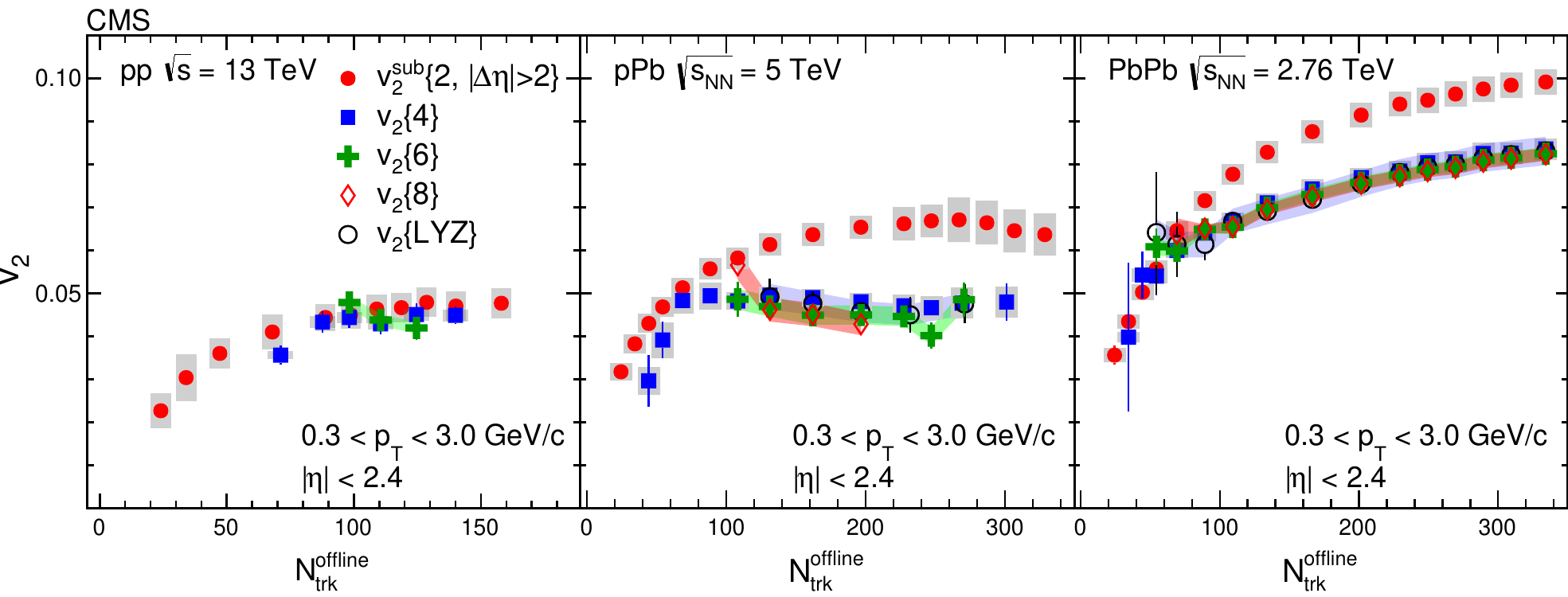}
  \caption{\label{fig:v24}
     Left: The $v_{2}^\text{sub}\{2, \abs{\Delta\eta}>2\}$, $v_{2}\{4\}$ and $v_{2}\{6\}$ values as a function
     of \noff for charged particles, averaged over $0.3 < \pt < 3.0\GeVc$
     and $\abs{\eta}<2.4$, in \pp collisions at $\roots = 13$\TeV.
     Middle: The $v_{2}^\text{sub}\{2, \abs{\Delta\eta}>2\}$, $v_{2}\{4\}$, $v_{2}\{6\}$, $v_{2}\{8\}$,
     and $v_{2}\{\mathrm{LYZ}\}$ values in \pPb collisions at $\rootsNN = 5$\TeV~\cite{Chatrchyan:2013nka}.
     Right: The $v_{2}^\text{sub}\{2, \abs{\Delta\eta}>2\}$, $v_{2}\{4\}$, $v_{2}\{6\}$, $v_{2}\{8\}$,
     and $v_{2}\{\mathrm{LYZ}\}$ values in \PbPb collisions at $\rootsNN = 2.76$\TeV~\cite{Chatrchyan:2013nka}.
     The error bars correspond to the statistical uncertainties, while the shaded areas denote the systematic uncertainties.
   }
\end{figure*}

To obtain $v_{2}\{4\}$ and $v_{2}\{6\}$ results using Eq.~(\ref{eq:v24}),
the cumulants are required to  be at least two standard deviations away from
their physics boundaries (\ie $c_{2}\{4\}/\sigma_{c_{2}\{4\}} < -2$ and
$c_{2}\{6\}/\sigma_{c_{2}\{6\}} > 2$), so that the statistical uncertainties can be
propagated as Gaussian fluctuations~\cite{Feldman:1997qc}. The $v_{2}\{4\}$
and $v_{2}\{6\}$ results, averaged over $0.3 < \pt < 3.0\GeVc$ and $\abs{\eta}<2.4$,
for \pp collisions at $\roots = 13$\TeV are shown in the left panel of Fig.~\ref{fig:v24},
as a function of event multiplicity. The $v_2$ data obtained from long-range
two-particle correlations after correcting for jet correlations ($v_{2}^\text{sub}\{2, \abs{\Delta\eta}>2\}$)
are also shown for comparison.

Within experimental uncertainties, the multi-particle cumulant
$v_{2}\{4\}$ and $v_{2}\{6\}$ values in high-multiplicity \pp collisions are
consistent with each other, similar to what was observed previously in
\pPb and \PbPb\ collisions~\cite{Khachatryan:2015waa}. This provides strong
evidence for the collective nature of the long-range correlations observed in \pp collisions.
However, unlike for \pPb and \PbPb collisions where
$v_{2}^\text{sub}\{2, \abs{\Delta\eta}>2\}$ values show a larger magnitude than multi-particle cumulant $v_2$ results,
the $v_2$ values obtained from two-, four-, and six-particle correlations are
comparable in \pp collisions at $\roots = 13$\TeV within uncertainties.
In the context of hydrodynamic models, the relative ratios of $v_2$ among two- and various
orders of multi-particle correlations provide insights to the details of initial-state
geometry fluctuations in \pp and \pPb systems. As shown in Ref.~\cite{Yan:2013laa},
the ratio of $v_{2}\{4\}$ to $v_{2}^\text{sub}\{2, \abs{\Delta\eta}>2\}$ is related to the total number of
fluctuating sources in the initial stage of a collision.
The comparable magnitudes of $v_{2}^\text{sub}\{2, \abs{\Delta\eta}>2\}$ and $v_{2}\{4\}$ signals
observed in \pp collisions, compared to \pPb collisions at similar multiplicities,
may indicate a smaller number of initial fluctuating sources that drive
the long-range correlations seen in the final state.
Meanwhile, it remains to be seen whether other
proposed mechanisms~\cite{Dusling:2015gta,Dusling:2012wy,Dusling:2012cg}
in interpreting the long-range correlations in \pPb and \PbPb\ collisions can also describe
the features of multi-particle correlations seen in \pp collisions.

\section{Summary}

The CMS detector has been used to measure two- and multi-particle azimuthal correlations
with $\PKzS$, $\PgL/\PagL$ and inclusive charged particles
over a broad pseudorapidity and transverse momentum range
in \pp collisions at $\roots = 5$, 7, and 13\TeV. With the implementation of high-multiplicity
triggers during the LHC 2010 and 2015 \pp runs, the correlation data
are explored over a broad particle multiplicity range. The observed long-range
($\abs{\deta}>2$) correlations are quantified in terms of azimuthal anisotropy
Fourier harmonics ($v_n$). The elliptic ($v_2$) and triangular ($v_3$) flow
Fourier harmonics are extracted from long-range two-particle correlations.
After subtracting contributions from back-to-back jet correlations estimated using
low-multiplicity data, the $v_2$ and $v_3$ values are found to increase with
multiplicity for $\noff \lesssim 100$, and reach a relatively constant value
at higher values of \noff. The \pt dependence of the $v_2$ harmonics in high-multiplicity \pp events
is found to have no or very weak dependence on the
collision energy. In low-multiplicity events, similar $v_2$ values as a function of
\pt are observed for inclusive charged particles, $\PKzS$ and \PgL/\PagL,
possibly reflecting a common back-to-back jet origin of the correlations for all particle species.
Moving to the higher-multiplicity region, a particle species dependence of $v_2$ is
observed with and without correcting for jet correlations. For $\pt \lesssim 2$\GeVc,
the $v_2$ of $\PKzS$ is found to be larger than that of \PgL/\PagL. This behavior%, which is consistent
is similar to what was previously observed for identified particles
produced in \pPb and \AonA collisions at RHIC and the LHC. This mass ordering
tends to reverse at higher \pt values. Finally, $v_2$ signals based on
four- and six-particle correlations are observed for the first time in \pp collisions. The $v_2$ values obtained with two-,
four-, and six-particle correlations at $\roots = 13$\TeV are found to be comparable within uncertainties.
These observations provide strong evidence supporting the interpretation of a collective origin for the observed long-range correlations in high-multiplicity pp collisions.

\begin{acknowledgments}
We congratulate our colleagues in the CERN accelerator departments for the excellent performance of the LHC and thank the technical and administrative staffs at CERN and at other CMS institutes for their contributions to the success of the CMS effort. In addition, we gratefully acknowledge the computing centers and personnel of the Worldwide LHC Computing Grid for delivering so effectively the computing infrastructure essential to our analyses. Finally, we acknowledge the enduring support for the construction and operation of the LHC and the CMS detector provided by the following funding agencies: BMWFW and FWF (Austria); FNRS and FWO (Belgium); CNPq, CAPES, FAPERJ, and FAPESP (Brazil); MES (Bulgaria); CERN; CAS, MoST, and NSFC (China); COLCIENCIAS (Colombia); MSES and CSF (Croatia); RPF (Cyprus); SENESCYT (Ecuador); MoER, ERC IUT and ERDF (Estonia); Academy of Finland, MEC, and HIP (Finland); CEA and CNRS/IN2P3 (France); BMBF, DFG, and HGF (Germany); GSRT (Greece); OTKA and NIH (Hungary); DAE and DST (India); IPM (Iran); SFI (Ireland); INFN (Italy); MSIP and NRF (Republic of Korea); LAS (Lithuania); MOE and UM (Malaysia); BUAP, CINVESTAV, CONACYT, LNS, SEP, and UASLP-FAI (Mexico); MBIE (New Zealand); PAEC (Pakistan); MSHE and NSC (Poland); FCT (Portugal); JINR (Dubna); MON, RosAtom, RAS and RFBR (Russia); MESTD (Serbia); SEIDI and CPAN (Spain); Swiss Funding Agencies (Switzerland); MST (Taipei); ThEPCenter, IPST, STAR and NSTDA (Thailand); TUBITAK and TAEK (Turkey); NASU and SFFR (Ukraine); STFC (United Kingdom); DOE and NSF (USA).

Individuals have received support from the Marie-Curie program and the European Research Council and EPLANET (European Union); the Leventis Foundation; the A. P. Sloan Foundation; the Alexander von Humboldt Foundation; the Belgian Federal Science Policy Office; the Fonds pour la Formation \`a la Recherche dans l'Industrie et dans l'Agriculture (FRIA-Belgium); the Agentschap voor Innovatie door Wetenschap en Technologie (IWT-Belgium); the Ministry of Education, Youth and Sports (MEYS) of the Czech Republic; the Council of Science and Industrial Research, India; the HOMING PLUS program of the Foundation for Polish Science, cofinanced from European Union, Regional Development Fund, the Mobility Plus program of the Ministry of Science and Higher Education, the OPUS program contract 2014/13/B/ST2/02543 and contract Sonata-bis DEC-2012/07/E/ST2/01406 of the National Science Center (Poland); the Thalis and Aristeia programs cofinanced by EU-ESF and the Greek NSRF; the National Priorities Research Program by Qatar National Research Fund; the Programa Clar\'in-COFUND del Principado de Asturias; the Rachadapisek Sompot Fund for Postdoctoral Fellowship, Chulalongkorn University and the Chulalongkorn Academic into Its 2nd Century Project Advancement Project (Thailand); and the Welch Foundation, contract C-1845.
\end{acknowledgments}

\clearpage
\bibliography{auto_generated}

\cleardoublepage \appendix\section{The CMS Collaboration \label{app:collab}}\begin{sloppypar}\hyphenpenalty=5000\widowpenalty=500\clubpenalty=5000\input{HIN-16-010-authorlist.tex}\end{sloppypar}
\end{document}

%% file: HIN-16-010-authorlist.tex
\textbf{Yerevan Physics Institute,  Yerevan,  Armenia}\\*[0pt]
V.~Khachatryan, A.M.~Sirunyan, A.~Tumasyan
\vskip\cmsinstskip
\textbf{Institut f\"{u}r Hochenergiephysik der OeAW,  Wien,  Austria}\\*[0pt]
W.~Adam, E.~Asilar, T.~Bergauer, J.~Brandstetter, E.~Brondolin, M.~Dragicevic, J.~Er\"{o}, M.~Flechl, M.~Friedl, R.~Fr\"{u}hwirth\cmsAuthorMark{1}, V.M.~Ghete, C.~Hartl, N.~H\"{o}rmann, J.~Hrubec, M.~Jeitler\cmsAuthorMark{1}, A.~K\"{o}nig, I.~Kr\"{a}tschmer, D.~Liko, T.~Matsushita, I.~Mikulec, D.~Rabady, N.~Rad, B.~Rahbaran, H.~Rohringer, J.~Schieck\cmsAuthorMark{1}, J.~Strauss, W.~Treberer-Treberspurg, W.~Waltenberger, C.-E.~Wulz\cmsAuthorMark{1}
% ...
\vskip\cmsinstskip
% ...
1:~~Also at Vienna University of Technology, Vienna, Austria\\
% ...
